\documentclass[11pt]{JHEP3}
\usepackage{amsmath}
\usepackage{amssymb}
\usepackage{bbold}
\usepackage{xspace}
\usepackage[numbers,sort]{natbib}
\usepackage[bottom]{footmisc}
\usepackage{graphicx}
\usepackage{psfrag}
\usepackage{relsize}
\usepackage[bf,small]{caption2}
\setcaptionwidth{.9\textwidth}
\DeclareMathOperator{\Imag}{Im}

\newcommand{\adss}{$\text{AdS}_5\times S^5$\xspace}
\newcommand{\dcite}[1]{{\citeauthor{#1}~\cite{#1}}}
\newcommand{\ddcite}[2]{{\citeauthor{#1}~\cite{#1,#2}}}
\makeatletter
\renewcommand\@ENVwarn[1]{}
\makeatother

\title{Holographic decays of large-spin mesons}
\author{Kasper Peeters{}$^1$, Jacob Sonnenschein{}$^2$ and Marija Zamaklar{}$^1$\\
\llap{{}$^1$}Max-Planck-Institut f\"ur Gravitationsphysik\\
Albert-Einstein-Institut\\ Am M\"uhlenberg 1\\ 14476 Golm, Germany\\
~\\
\llap{{}$^2$}School of Physics and Astronomy\\
The Raymond and Beverly Sackler Faculty of Exact Sciences\\
Tel Aviv University\\
Ramat Aviv, 69978, Israel\\
~\\
\email{kasper.peeters@aei.mpg.de}\\
\email{cobi@post.tau.ac.il}\\
\email{marija.zamaklar@aei.mpg.de}}

\abstract{We study the decay process of large-spin mesons in the
  context of the gauge/string duality, using generic properties of
  confining backgrounds and systems with flavour branes. In the string
  picture, meson decay corresponds to the quantum-mechanical process
  in which a string rotating on the IR~``wall'' fluctuates, touches a flavour
  brane and splits into two smaller strings. This process
  automatically encodes flavour conservation as well as the Zweig
  rule. We show that the decay width computed in the string picture is
  in remarkable agreement with the decay width obtained using the
  phenomenological Lund model.}

\keywords{AdS/CFT, meson decay, spinning strings}
\preprint{hep-th/0511044\\AEI-2005-150\\TAUP-2815/05}
\begin{document}
\section{Introduction}

Understanding the gauge/string correspondence in the context of
realistic, non-supersym\-met\-ric, confining gauge theories remains a
major open problem. No fully satisfactory geometries dual to confining
gauge theories are known so far, and there are general arguments that,
in order to fully describe QCD, one will have to go beyond simple
supergravity considerations. Nevertheless, it is quite remarkable that
many qualitative properties of confining theories \emph{do} get
reproduced correctly from computations in dual supergravity
theories. So far, a large body of work in this field has been
concerned with a comparison of hadron spectra with the spectra of
states on the gravity side. This is a rather kinematical test, and one
wonders whether more dynamical properties, such as decay rates, may
perhaps also be captured by the correspondence. In recent work
by~\dcite{Sakai:2004cn} decays of \emph{low-spin} particles (which are
captured by the supergravity and DBI modes) have been considered.  The
decays of \emph{high-spin} mesons, which correspond to genuine stringy
processes, have, however, not been addressed so far.  In the present
paper, we initiate the study of high-spin meson decays using the dual
string theory description.

\begin{figure}[t]
\begin{center}
\psfrag{lm}{\vbox{\hbox{\small large mass}\hbox{\small flavour brane}}}
\psfrag{im}{\vbox{\hbox{\small intermediate mass}\hbox{\small flavour brane}}}
\psfrag{wl}{\hbox{\small infrared ``wall''}}
\includegraphics[width=.65\textwidth]{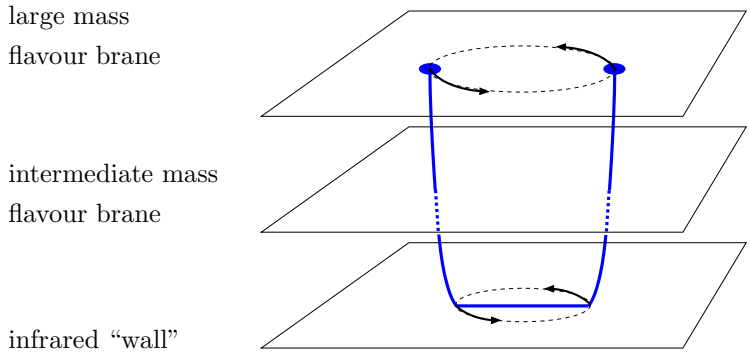}
\end{center}
\caption{A high-spin meson composed of heavy quarks, represented in
  the dual string picture as an open string ending on a flavour brane
  far away from the infrared ``wall''. To good approximation, the string
  consists of two vertical segments, called ``region I'' and one
  horizontal segment, called ``region II''.\label{f:Ustring}}
\end{figure}

In the context of gauge/string duality, mesons are incorporated by
adding one or more flavour branes into confining dual
geometries~\cite{Karch:2002sh}.  In this setup, mesons of low-spin are
identified with small fluctuations of the flavour brane. This has
allowed for a computation of masses and decay rates, both in the
approximation in which the flavour brane is treated as a
probe~\cite{Sakai:2004cn,Sakai:2005yt} and in the case where the
backreaction is taken into account~\cite{Erdmenger:2004dk}. However,
the supergravity (i.e.~DBI) approximation is not sufficient to deal
with mesons with spin larger than one, which from a phenomenological
point of view are at least as important.  These mesons are described
by genuine string excitations, and their precise treatment would
require the quantisation of strings in the confining geometries. Given
the complexity of the candidate dual geometries, this is a rather
formidable task. However, very high-spin mesons can be described in
this set up using \emph{semi-classical} macroscopic spinning string
configurations.

The open string configuration which we will consider is depicted in
figure~\ref{f:Ustring}.  This is a U-shaped string, with its endpoints
located on the flavour branes, and which is pulled towards the
infrared ``wall'' by the gravitational potential. It also extends along
the ``wall'', where it is prevented from collapse by its rigid rotation.
This string is equivalent to a system two quarks connected by a flux
tube, with masses proportional to the distance from the flavour branes
to the ``wall''.  Thus one can have high-spin mesons with light, medium or
heavy quarks.  The spectrum exhibits deviation from Regge behaviour
with appropriate non-linear corrections which depend on the quark
masses~\cite{PandoZayas:2003yb,Kruczenski:2004me}.

\begin{figure}[t]
\begin{center}
\psfrag{flavour brane}{\smaller\smaller flavour brane}
\psfrag{IR wall}{\smaller\smaller infrared ``wall''}
\includegraphics*[width=.8\textwidth]{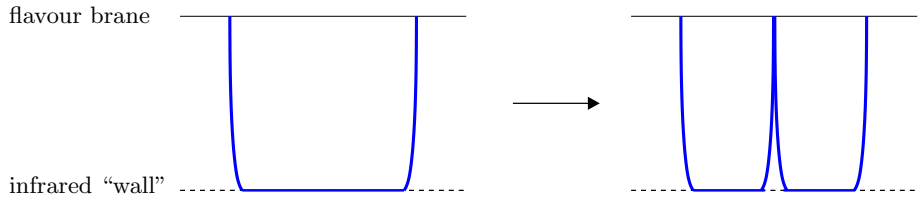}
\end{center}
\caption{The basic idea behind the description of high-spin mesons in
  duals of confining gauge theories (left). The open string
  corresponding to the meson starts on a flavour brane, stretches to
  the infrared ``wall'', and then reaches up again to the (same or
  another) flavour brane. A decay process (right) requires that the
  string fluctuates, touches the flavour
  brane and then reconnects to it.\label{f:basicsetup}}
\end{figure}

In the present paper we study decays of high-spin mesons in detail,
both from a qualitative as well as from a quantitative point of view.
Semi-classically, the U-shaped string can decay due to an instability
of its endpoints or due to breaking of the string itself. The first
type of decay channel is associated to radiation processes on the
flavour brane, and will be discussed a separate publication.  To
describe the second family of decay channels, recall that an open
string always has to end on a brane, and that therefore the string can
break semi-classically if and only if one (or more) of its middle
points touch one of the flavour branes.  If no flavour brane is
present on the infrared ``wall'', then classically this condition is
never satisfied for the U-shaped string of figure~\ref{f:Ustring},
except at those points where the vertical part of the string
intersects a flavour brane (splitting at these points, i.e.~of the
vertical segments of the string, was analysed in~\cite{Cotrone:2005fr}
and turns out to be highly suppressed). However, semi-classically, there
is a finite probability that the string, due to the quantum
fluctuations, touches one of the flavour branes, splits and gets
reconnected to it, producing two or more outgoing mesons
(i.e.~``hanging'' open strings, see figure~\ref{f:basicsetup}).

Therefore, in order to compute the decay rate semi-classically, we
need to compute the probability of the horizontal part of the string
to touch a flavour brane and the probability that the string splits
when it is on the brane.\footnote{Fluctuations of the vertical parts
of the strings are also possible, and would lead to decay channels in
which the initial meson decays into a meson (i.e.~a hanging open
string) and a glue ball (i.e.~a closed string). These channels are
more suppressed due to the centrifugal force which suppresses the
transverse fluctuations, and additional powers of~$g_s$ which suppress
open-to-closed string amplitudes with respect to open-to-open string
amplitudes. We will therefore not discuss these processes.}  Although
the calculation of the string fluctuation probability is a hard task,
we have found several simplification and approximation methods which
make it feasible. The main idea is to focus on the part of the
geometry near the ``wall'', and then construct the string wave function in
this simplified geometry by semi-classical quantisation (for details
see section~\ref{s:decayrates}).  Once this is achieved, the
probability for finding the string at a certain distance from the ``wall''
can be extracted.  On the other hand, for a string which is located on
the brane, the probability for it to split at any given point can be
computed using the flat space results of Dai and
Polchinski~\cite{Dai:1989cp} and Mitchell and
Turok~\cite{Mitchell:1987th}.  We expect that these semi-classical
computations should capture the main features of the full string decay
process. That this is indeed the case can be shown explicitly in flat
space, by comparing this splitting rate with a full quantum
computation.

In gauge theory (i.e.~QCD), meson decay widths are not easily
computable from first principles because of strong coupling
problems. A heuristic model has therefore been developed some thirty
years ago, which goes under the name of the ``Lund'' model, and very
successfully describes decays of various mesons.  In this model a
meson is described by two massive particles (quarks) connected by a
massless relativistic string which models the strong force between the
quarks. The probability that a string splitting event occurs is
determined by the Schwinger pair production probability. This model is
in widespread use in event generators such as
Pythia~\cite{Sjostrand:2003wg}, and has turned out to be surprisingly
effective. We will see that the main qualitative features of the Lund
model are, indeed, reproduced from the holographic stringy dual
computation (for a comparison between our model and the Lund model,
see section~\ref{s:summary}). Moreover, properties such as the Zweig
rule, which have to be ``added by hand'' to the Lund model, are an
automatic consequence of the holographic description. Our model also
predicts some deviations from the Lund model for very large values of
the spin, but there is unfortunately not yet enough experimental data
in this regime to see whether those corrections are indeed required.

In order to make this paper self-contained, we first review in
section~\ref{s:review} in some detail the dual picture of mesons, as
it arises in the string/gauge theory correspondence (readers familiar
with this material can safely skip to the next section). In
section~\ref{s:qualitative} we then give a qualitative description of
the decay process of mesons, both in the old phenomenological models
as well as in the new setup. Our main quantitative result is presented
in section~\ref{s:decayrates}, where we show that the decay rates
computed in the new picture indeed agree with the rates obtained in
the Lund model.  The reader who is not interested in any technical
details, but rather in our setup and in comparison with experiment, is
advised to go directly to section~\ref{s:summary}, which can be read
independently.

%%%%%%%%%%%%%%%%%%%%%%%%%%%%%%%%%%%%%%%%%%%%%%%%%%%%%%%%%%%%%%%%%%%%%%%

\section{A review of the dual picture of  mesons}
\label{s:review}
\subsection{Supergravity duals of confining gauge theories}

To construct the holographic picture of mesons one first has to
specify a supergravity model which is dual to the desired confining
gauge theory.  By now, there are several supergravity backgrounds
which are know to be associated to confining gauge dynamics.  An
important model based on near-extremal~D4-branes~\cite{Itzhaki:1998dd}
was shown to exhibit, in the limit of large temperature, features of
the low energy regime of strongly coupled pure Yang-Mills
theory~\cite{Witten:1998zw}. In particular, the Wilson loop of these
models was shown to exhibit an area law behaviour, as required by
a confining theory~\cite{Brandhuber:1998er}.  Recently, an analogous
non-critical supergravity model was proposed, which is dual to the
same gauge system, but without a contaminating Kaluza-Klein
sector~\cite{Kuperstein:2004yk}.

In this work we study a mechanism for meson decays which does not rely
heavily on the details of the confining background, but rather uses
generic features which all known confining geometries
possess. However, some explicit parts of the computations will be
performed for a prototype class of models based on near-extremal
D4-branes. Therefore, let us first briefly review the main features of
this model. The basic setup is that of \mbox{type-IIA} superstring
theory with a set of~$N_c$ D4-branes that wrap a
circle~\cite{Witten:1998zw}. Fermions are taken to have anti-periodic
boundary conditions along the circle.  The corresponding near-horizon
limit of the background consists of a metric, a running dilaton, and a
four-form RR field strength given by
\begin{equation}
\label{D4}
\begin{aligned}
{\rm d}s^2 &= \left(\frac{U}{R} \right)^{3/2} \big[ \eta_{\mu\nu}
  {\rm d} X^{\mu} {\rm d} X^{\nu} + f(U)  {\rm d} \theta^2 \big] + 
\left(\frac{R}{U}\right )^{3/2}\left [\frac{{\rm d}U^2 }{f(U)} + U^2 {\rm d}\Omega_4 \right ]  \\[1ex]
e^\phi &= g_s \left(\frac{U}{R} \right)^{3/4}\,, \qquad F_4=\frac
  {2\pi N_c}{V_4}\epsilon_4 \,,
\qquad  f(U) = 1 - \left(\frac{U_\Lambda}{U}\right)^3 \,.
\end{aligned}
\end{equation}
Here $U$ is the radial direction, which has dimension of length and is
bounded from below by~$U\geq U_\Lambda$. We will refer to $U
=U_\Lambda$ as the ``wall'' of space-time. Note, however, that this is
only a wall in coordinate space, in the same sense in which $r=0$ in
polar coordinates is a ``wall'' of the plane. The geometry
near~$U=U_\Lambda$ is actually cigar-like. Extending
beyond~$U=U_\Lambda$, one ends up on the other side (i.e.~at an
antipodal point) of the cigar.\footnote{For future reference, let us
also recall that the four-dimensional Yang-Mills gauge coupling is
related to the other parameters by
\begin{equation}
g_{\text{YM}}^2 N = 2\,M_\Lambda\,R^3 / \alpha' \,;
\end{equation}
details of the derivation of this formula and the other expressions in
this section can be found in~\cite{Kruczenski:2004me}.}

The worldvolume coordinates of the D4-branes are along the $X^{\mu}$
$(\mu=0,1,2,3)$ directions and~$\theta$ is the thermal circle.  The
line element of the unit four-sphere is denoted by~${\rm d}\Omega_4$,
its volume by $V_4=\tfrac{8}{3}\pi^2$, its volume-form by $\epsilon_4$
and its radius~$R$ is given by $R^3= \pi g_s N_c l_s^3$ where $l_s$ is
the string length.  The size of the thermal circle follows from the
requirement that the metric does not have a conical singularity on the
``horizon'' at $U=U_\Lambda$, and is given by
\begin{equation}
\label{E0}
L_{\Lambda} =  \frac{4}{3} \pi \left( \frac{R^3}{U_\Lambda} \right)^{1/2} \, .
\end{equation}
It is important to note that the mass scale~$M_\Lambda =
2\pi/L_\Lambda$ is also the scale of lowest lying Kaluza-Klein
excitation and hence that the theory appears to be four dimensional if
probed below the energy scale
\begin{equation}
\label{E1}
E < \frac{3}{2} \left( \frac{U_\Lambda}{R^3} \right)^{1/2} \, .
\end{equation}
The supergravity regime is valid (i.e.~one can forget about
higher-derivative corrections) if the curvature radius is larger
than~$\sqrt{\alpha'}$,
\begin{equation}
\label{E2}
{\mathcal R}^4 \equiv U_\Lambda R^3 \gg \alpha'^2 \, .
\end{equation}
Finally, the condition that string theory is perturbative (at the
wall) requires that
\begin{equation}
\label{E3} 
e^{\phi} < 1\quad \Rightarrow\quad  g_s < \left( \frac{U_ \Lambda}{R} \right)^{-3/4}\, .
\end{equation}
In summary, the prototype geometry~\eqref{D4} exhibits all features
which are \emph{generic} for confining geometries, and will be
necessary for our generic considerations of meson decays. The space
caps off at some distance in the radial direction (corresponding to
the confining energy scale). At every fixed radial slice, the space
has four-dimensional Lorentz invariance (corresponding to the
directions parallel to the ``wall'') times the internal direction
(corresponding the required global symmetries of the theory, but
giving rise to unwanted Kaluza-Klein excitations). And the warping of
the space in the transverse direction is such that the Wilson loop in
this geometry shows area law behaviour.

\subsection{Flavour branes in confining backgrounds}
\label{s:flavour_branes}

To describe the holographic picture of mesons requires the
introduction of additional flavour branes to the system of branes
which give rise to confining geometries.  If the number of flavour
branes is small enough, these can be treated as probes, whose dynamics
is governed by the DBI action.  The open strings between the
original~$N_c$ branes and the flavour branes play the role of quarks
(anti-quarks) in the fundamental (anti-fundamental) representation of
the colour and flavour groups.  This way of incorporating fundamental
quarks was originally proposed by~\dcite{Karch:2002sh} in the context
of the $\text{AdS}_5\times S^5$ model.  The first application of these
ideas to a confining model was made in~\cite{Sakai:2003wu} with
D7-branes probing the Klebanov-Strassler geometry. Bending of the
probe brane due to the gravitational potential is an important effect,
as it was shown to be associated to $U(1)_A$~symmetry breaking in the
work of~\cite{Babington:2003vm}.  Flavour D6-branes were introduced
into the model given by~\eqref{D4}
in~\cite{Kruczenski:2003uq}.\footnote{Flavour branes have since been
introduced in many other confining and non-confining models (see
e.g.~\cite{Nastase:2003dd,Wang:2003yc,Nunez:2003cf,Ouyang:2003df,Kruczenski:2003be,Hong:2003jm,Evans:2004ia,Bando:2004ct,Ghoroku:2004sp,Erdmenger:2004dk,Arean:2004mm,Apreda:2005hj}).
Other related holographic models for hadrons have also been
built~\cite{Son:2003et,Erlich:2005qh,DaRold:2005zs,deTeramond:2005su},
achieving notable success for their quantitative predictions.}  In
order to exhibit flavour chiral symmetry breaking, one has to consider
models which exhibit~\mbox{$U_L(N_f)\times U_R(N_f)$} chiral symmetry.
The recent model of~\ddcite{Sakai:2004cn}{Sakai:2005yt} incorporates
this phenomenon by introducing D8/$\overline{\text{D8}}$-branes as
flavour probe branes.  An analogous non-critical model based on
D4/$\overline{\text{D4}}$-branes was analysed in~\cite{Casero:2005se}.
For the purpose of our calculations, the distinction between all these
models is, however, irrelevant. Because the spinning string
configurations are readily available for the model with
D6-branes~\cite{Kruczenski:2003uq}, we will restrict to this case in
our explicit computations.

\begin{figure}[t]
\begin{center}
\psfrag{D4}{D4}
\psfrag{D6}{D6}
\psfrag{r}{$r$}
\psfrag{ri}{$r_{\infty}$}
\psfrag{l}{$\lambda$}
\psfrag{rf}{$\rho_f$}
\psfrag{rl}{$\rho_{\Lambda}$}
\includegraphics[width=.6\textwidth]{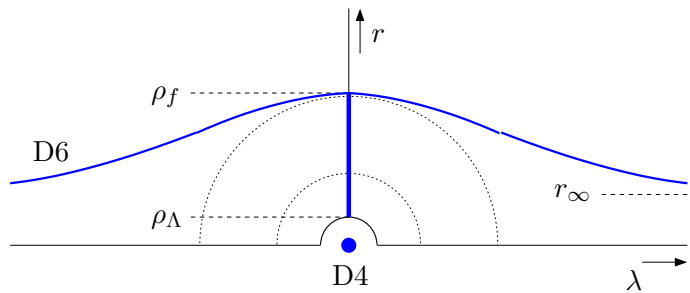}\vspace{-8ex}
\end{center}
\caption{Schematic overview of the embedding of the probe D6-brane,
  described by~$r(\lambda)$, into the geometry of the stack of
  D4-branes (negative values of~$\lambda$ correspond to points with
  $\phi\rightarrow \phi+\pi$ while negative values of~$r$ correspond
  to~$\theta\rightarrow \theta+\pi$). The dotted half-circles are
  equal-potential lines of the gravitational field, the solid
  half-circle is the IR ``wall''. Also depicted is a high-spin meson,
  represented by the thick vertical line. This is a side-on view of an
  open string stretching from the flavour D6-brane to the ``wall'', along
  the ``wall'', and then back up to the flavour D6-brane.\label{f:D4D6}}
\end{figure}

In order to describe flavour probe D6-branes in the
geometry~(\ref{D4}), it is more convenient to introduce different
coordinates in the metric in the directions transverse to the
``wall''. A metric adapted to the embedding of the D6-brane
is~\cite{Kruczenski:2003uq}
\begin{equation}
\begin{aligned}
{\rm d}\tilde{s}^2 &=  
  K(\rho) \Big[ {\rm d}\rho^2 + \rho^2 {\rm d}\Omega_4^2\Big] 
= K(\rho) \Big[ {\rm d}\lambda^2 + \lambda^2 {\rm d}\Omega_2 + {\rm
  d}r^2 + r^2 {\rm d}\phi^2 \Big]\,, \\[1ex]
U(\rho)^{3/2} &\equiv \rho^{3/2} + \frac{U_\Lambda^2}{4 \rho^{3/2}}
  \,, 
 \qquad K(\rho) = R^{3/2} U^{1/2} \rho^{-2}\,, \qquad
\rho^2=\lambda^2 + r^2 \, . \\
\end{aligned}
\end{equation}
The probe D6-brane extends in the ``wall'' directions, fills out the
$S^2$ sphere spanned by~${\rm d}\Omega_2$ and has nontrivial profile
in the $r,\lambda$ plane.  The equation for the brane profile
$r(\lambda)$ follows from the DBI action, and can be solved
approximately in various regions.  The shape of the D6-brane in these
direction is depicted in figure~\ref{f:D4D6}: it stretches in the $r$
direction from~$r=\rho_{f}$ at $\lambda=0$ to $r=r_{\infty}$ at
$\lambda\to \infty$. Due to the non-trivial profile of the D6-brane,
the $U(1)_A$ symmetry (corresponding to rotations in the~$r,\phi$
plane, i.e.~in the~$\phi$ direction) is spontaneously broken and the
quark condensate $\langle \bar q q\rangle$ is non-zero. Asymptotically
one has $r= r_{\infty}+\frac{c}{\lambda}$ where $r_{\infty}$ is
related to the QCD (current algebra) quark mass and $c$ is related to
the quark condensate. The $U(1)_A$ symmetry is thus restored
asymptotically when the quark mass is set to zero, but spontaneously
broken due to the bending of the brane. Equipped with this information
we now proceed to describe the mesons of these models.

%%%%%%%%%%%%%%%%%%%%%%%%%%%%%%%%%%%%%%%%%%%%%%%%%%%%%%%%%%%%%%%%%%%%%%%

\subsection{High-spin mesons} 
\label{s:highspinmesons}

The spectrum of the (pseudo) scalar and vector mesons can be extracted
in the dual supergravity backgrounds from the spectrum of the
fluctuation of the flavour branes. Just as for glueballs, these can
only account for the meson states with spin smaller than or equal to
one. The other mesons should be captured by genuine string
excitations, which are generically very hard to analyse. However, when
the spin of the string becomes very large, further simplifications
occur, and classical solutions of the string sigma model can be used.

A particularly interesting large, open string configuration, was
recently constructed by~\dcite{Kruczenski:2004me}
and~\dcite{Paredes:2004is}.  This is an open, U-shaped string as
depicted in figure~\ref{f:Ustring} and~\ref{f:D4D6}. It hangs from the
probe D6-brane and is pulled by the gravitational force towards the
``wall'' of the background. At the same time, the string rotates in the
plane parallel to the ``wall'', and is extended in this direction due to
the centrifugal force.  More precisely, the region spanned by the open
string can be divided into two parts:
\begin{itemize}
\item Region I,  a ``vertical''  section  characterised by
  $\displaystyle\frac{{\rm d}\rho}{{\rm d} R}\rightarrow \infty$\,, 
\item Region II,  a ``horizontal'' section $\displaystyle\frac{{\rm
	 d}\rho}{{\rm d} R}\rightarrow 0 $\,, 
\end{itemize} 
where $R^2 = (X^1)^2 + (X^2)^2$ and $X^1, X^2$ is the plane of
rotation of the string.  In the limit of large angular momentum and
hence large separation, the string is indeed well-approximated by two
vertical segments and one horizontal one, and explicit simple
solutions can be found separately in these two regions.

It was further realised in~\cite{Kruczenski:2004me} that this
classical string configuration can be viewed as a rigid open string
with two massive endpoints, where the masses of the particles are
proportional to the vertical parts of string. The equivalence comes
about as follows.  To ``sew'' the solutions in regions~I and~II one
has to impose the condition that the string endpoints move with the
same velocity as the vertical parts of the string,
\begin{equation} 
1-\omega^2 L^2=\omega^2  
L \frac{1}{(U_\Lambda/R)^{3/2}}\int_{\rho_{\Lambda}}^{\rho_f}
{\rm d}\rho \frac {U(\rho)}\rho = \omega^2L \frac{m_q}{T_{\text{eff}}}\,,
\label{cond} 
\end{equation} 
where on the right-hand side of the equation we have used the
expression for the mass of the dynamical quarks~\cite{Kinar:1998vq},
\begin{equation} 
\label{mq} 
m_q=\frac{1}{2\pi\alpha'}\int_{\rho_f}^{\rho_\Lambda}\! {\rm d}\rho\,\sqrt{g_{00}g_{\rho\rho}}= 
\frac{1}{2\pi\alpha'}\int_{\rho_f}^{\rho_\Lambda}\!{\rm d}\rho\,\frac{U}{\rho}\,.
\end{equation} 
There are several arguments in favour of identifying the mass of the
vertical part of the string with the constituent mass of the quark,
and not a current algebra mass.

The relation~\eqref{cond} is precisely the relation that one derives
for a string with two massive endpoints of mass~$m_q$.  Indeed by
evaluating the energy and angular momentum of the string in the two
regions one finds that
\begin{equation}
\begin{aligned}
E_{\text{I}}&=\frac{2m_q}{\sqrt{1-\omega^2 L^2}}\,, \qquad &
J_{\text{I}}&=\frac{2m_q \omega L{}^2}{\sqrt{1-\omega^2 L^2}} \,,\\[1ex]
E_{\text{II}} &=
  T_{\text{eff}}\,\frac{2}{\omega}\arcsin(\omega
  L)\,,\quad&
J_{\text{II}} &=
  T_{\text{eff}}\,
 \frac{1}{\omega^2}\left(\arcsin(\omega L)-\omega L\sqrt{1-\omega^2
	L{}^2}\right)\,.
\end{aligned}
\end{equation}
The expressions in region~I are those for two spinning relativistic
particles. In region~II we find the energy and angular momentum of an
open string in flat spacetime, with an effective string
tension~$T_{\text{eff}}$, given by
\begin{equation}
\label{e:Teff}
T_{\text{eff}}=  \frac{1}{2 \pi \alpha'}
\sqrt{g_{00} g_{xx}\phantom{\big|}}\big|_{\text{wall}}
= \frac{1}{2 \pi \alpha'} \left(\frac{ U_{\Lambda}}{R} \right)^{3/2}
=  \frac{2}{27\,\pi}M_\Lambda^2\, (g^2_{\text{YM}} N)\,.
\end{equation}
Combining the results from the two regions we get
\begin{equation}
\label{EJ}
E = 2 \frac{T_{\text{eff}}}{\omega} \left( \arcsin \, x + \sqrt{\frac{m_q}{T L}} \right)\,, \quad
J = \frac{T_{\text{eff}}}{\omega^2} \left( \arcsin \, x +  x^2 \, \sqrt{\frac{m_q}{T L}} \right) \,,
\end{equation}
% These expressions are corrected versions of equations
% (3.8) or (4.29) in hep-th/0410035. 
where $x \equiv \omega L$. 
For fixed~$m_q$ and $T_{\text{eff}}$, there is only
one free parameter, for example~$L$, which uniquely fixes all other
parameters: the energy~$E$, the angular momentum~$J$ and the angular
velocity~$\omega$.

There are two important limits of this solution which will be relevant
for us later. The first limit is the one in which the dynamical quarks
are very light. This limit is relativistic, as the velocity~$x$ of the
endpoints tends to the velocity of light.  The energy and angular
momentum reduce to
\begin{equation}
\label{EJ-low}
x\rightarrow 1:\quad\quad E \rightarrow \pi T_{\text{eff}} L \, ,\, \, \quad  
J \rightarrow \frac{\pi}{2} T_{\text{eff}} L^2  \quad\Rightarrow\quad
J \rightarrow \frac{1}{2\pi T_{\text{eff}}}E^2 \, ,
\end{equation}
% See the cobi_3.8.nb notebook
i.e.~we recover the standard Regge trajectory in flat space with the
effective tension (\ref{e:Teff}), as expected. We thus see that the mass
of the U-shaped high-spin mesons is of the order
\begin{equation}
\label{hm}
M_{\text{high}} \sim M_{\Lambda} \sqrt{g_{YM}^2 N} \, 
\end{equation}
while recall that masses of the low spin (supergravity) mesons were
$M_{\text{low}} \sim M_{\Lambda} \sim M_{KK}$. We thus see that in the
supergravity regime, where $g_{YM}^2 N \gg 1$, there is a gap between
the low and high-spin mesons which hints at the fact that the
holographic dual of hadron physics will require $g^2_{\text{YM}} N\sim
1$.

The second limit is the limit of heavy quarks, i.e.~the
non-relativistic limit
\begin{equation}
\label{e:heavy_mq}
x\rightarrow 0: \quad\quad  m_q \approx \frac{T_{\text{eff}}  L}{x^2} \rightarrow \infty
\quad \Rightarrow \quad m_q \gg T_{\text{eff}} L \,. 
\end{equation}  
% See the cobi_3.8.nb notebook
This in turn implies
\begin{equation}
\label{EJ-high}
\begin{aligned}
E &= 2 T_{\text{eff}}\, L \bigg(1 + \frac{1}{x^2} + {\mathcal O}(x^3)\bigg) &&\rightarrow 2\, m_q + T_{\text{eff}}\, L\,, \\ 
J &= 2 T_{\text{eff}}\, L \bigg( \frac{1}{x} + {\mathcal O}(x) \bigg)  &&\rightarrow  2\sqrt{T_{\text{eff}} m_q}\, L^{3/2} \, .
\end{aligned}
\end{equation}
We see that in this limit, the energy and angular momentum blow up as
one would expect: it takes an infinite amount of energy to spin very
heavy particles. Note that in both these limits, whether the quark is
light or heavy is measured with respect to the total mass of the flat
part of the string. This mass is given by~$T_{\text{eff}} L$, rather
than by~$T_s L$,
\begin{equation} 
m_q  = - 2 T_{\text{eff}}\, L \, \delta x \, \,  \Leftrightarrow  \, \, m_q \ll  2 T_{\text{eff}}\, L \, .
\end{equation}
The relations~\eqref{EJ-high} imply that for a fixed and finite
energy, the length of the string has to go to zero (in units of
$1/\sqrt{T_{\text{eff}}}$) as the mass of the quarks is increased .

It is also straightforward to generalise the expressions of the energy
and angular momentum of the classical meson~\eqref{EJ} to the case of
a meson composed of quarks of two different masses; details can be found
in~\cite{Wilczek:2004im}. In general one can associate a different
value of~$\rho_f$ to each of the stacks of the probe brane,
namely~$\rho_{f_i}$ to the~$i^{\text{th}}$ stack.  In presently
available holographic setups, there are no limitations on the
locations~$\rho_{f_i}$ and the corresponding quark masses. For
convenience we group them into three classes according to the value of
distance from the ``wall'', which translates to three types of quark
masses:
\begin{subequations}
\begin{align}
\label{light}
m_l &\approx T_{\text{eff}}\,(\rho_{f_l}-\rho_{\Lambda}) \ll \Lambda_{\text{QCD}}\,, \\[1ex]
\label{midi}
m_m &\approx T_{\text{eff}}\,(\rho_{f_l}-\rho_{\Lambda}) \sim  \Lambda_{\text{QCD}}\,, \\[1ex]
\label{heavy}
m_h &\approx  T_{\text{eff}}\,(\rho_{f_l}-\rho_{\Lambda}) \gg \Lambda_{\text{QCD}}\,. 
\end{align}
\end{subequations}
Accordingly, there are six classes of mesons according to the possible
different probe branes on which the stringy meson ends,
\begin{equation}
(l, l)\, ,\, \, (l, m)\, , \,\,  (l, h)\, , \, \, (m , m)\, , \, \,  (m, h) \, ,\, \, (h, h) \, . 
\end{equation}

%%%%%%%%%%%%%%%%%%%%%%%%%%%%%%%%%%%%%%%%%%%%%%%%%%%%%%%%%%%%%%%%%%%%%%%

\section{Old and new descriptions of meson decay}
\label{s:qualitative}
\subsection{The Casher-Neuberger-Nussinov model}

Having reviewed the dual kinematical picture of glueballs and mesons,
we now want to focus on dynamical aspects. In the present section we
will compare the qualitative aspects of the old, phenomenological
picture of meson decay with the new picture as it arises from the
gauge/string correspondence. A quantitative discussion follows in
section~\ref{s:decayrates}.

In \cite{Casher:1978wy} the decay of a meson, or rather the process of
multiple quark pair production, was described in terms of a model
where a meson is built from a quark/anti-quark pair with a colour
electric flux tube between them.\footnote{This model was also
suggested independently at around the same time
by~\dcite{Gurvich:1979nq}, who obtained similar qualitative results
as~\dcite{Casher:1978wy}.} When a new pair is created at a certain
point along the flux tube, it will be pulled apart and tear the
original tube into two tubes.
The model is based on two assumptions: (i)~that at the hadronic energy
scale of 1~GeV the quarks can be treated as Dirac particles with
constituent masses; (ii)~that there is a chromo-electric flux tube of
universal thickness which is being created in a timescale that is
short compared to the hadronic timescale. The chromoelectric field is
treated as a classical longitudinal abelian field.\footnote{A more
precise calculation, which does not rely on the WKB approximation and
probes the full non-abelian structure of the flux tube, was recently
presented by~\dcite{Nayak:2005pf}.} The flux tube is parametrised by
the radius of the tube~$r_t$, the gauge coupling~$g$ which is also the
charge of the quark and the electric field~${\mathcal E}_t$. The
energy per unit length stored in the tube is equal to the string
tension,
\begin{equation}
T_{\text{eff}}=\frac{1}{2} {\cal E}_t^2\pi\, r_t^2= \frac{1}{2\pi \alpha'}= \frac{1}{4} g\, {\cal E}_t
\end{equation}
where in the last part of the equation the Gauss law was used.  It is
easy to verify that $g^2=4r_t^2/\alpha'$. When the radius of the flux
tube is smaller than the size of the tube but larger than the distance
scale relevant to pair production, i.e.~when it is of the order of
$r_t\sim 2.5~\text{GeV}^{-1}$, the coupling constant is indeed weak,
$g^2/8\pi<1$.

The process of pair creation inside the tube is mapped to a system of
Dirac particles of mass~$m_q$ interacting with a constant electric
field, which was solved by Schwinger. From the probability of a single
pair-creation event to occur,
\begin{equation}
\label{e:singledecay}
P_{\text{pair prod.}} = \exp\left( - \frac{\pi m_q^2}{2\,T_{\text{eff}}}\right)\,,
\end{equation}
one derives~\cite{Casher:1978wy} the decay probability per unit time
and per unit volume,
\begin{equation}\label{decaypro}
P= \frac{g^2 {\cal E}^2 }{16\pi^3} 
  \sum_{n=1}^\infty \frac{1}{n^2} \exp\left(-\frac{2\pi m_q^2 n }{g{\cal E}}\right)
  =  
  \frac{T_{\text{eff}}^2}{\pi^3} \sum_{n=1}^\infty \frac{1}{n^2} \exp\left(-\frac{\pi m_q^2 n }{2 T_{\text{eff}}}\right)\,.
\end{equation}   
This probability was then used to determine the probability of a meson
to decay, which was found to be $1- e^{-V_4(t_M)P}$ where~$t_M$ is the
meson lifetime measured in its rest frame. The volume of the system
was computed for a rotating flux tube and for a one dimensional
oscillator. For the first case we use $M =\pi T_{\text{eff}} L$ which
implies that $V_4(t_M)= \pi r_t^2 L t_M$ and hence the decay width is
$\Gamma= \pi r_t^2 L P$ and finally
\begin{equation}
\left(\frac{\Gamma}{M}\right)_{\text{rot}} = \frac{2 r_t^2}{T_{\text{eff}}}P = (0.6-8.5)\times 10^{-2} \,.
\end{equation} 
The numerical value was derived by using~\eqref{decaypro} for the
decay probability, introducing constituent masses for $m_u$, $m_d$ and
$m_s$ (taken to be 75~MeV for the light quarks and 400~MeV for the
heavy ones, with~$T_{\text{eff}}\approx 0.177~\text{GeV}^2$) and
summing it over the three flavours.
 
For the case of the oscillator, the relation between the length and
the mass is given by $M= T_{\text{eff}} L$ and therefore on average
$V_4(t_M)= \tfrac{1}{2} \pi r_t^2 L t_M$ and hence $\Gamma=\frac{1}{2}
\pi r_t^2 L P$ which means that
\begin{equation}
\left(\frac{\Gamma}{M}\right)_{\text{osc}}=\frac{\pi}{4} \left(\frac{\Gamma}{M}\right)_{\text{rot}}\,.
\end{equation} 
From expression~\eqref{decaypro} two properties are immediately clear: the
exponential suppression does not depend on the length of the string,
while the total probability scales linearly in the length.

Let us finally also mention that the Casher-Neuberger-Nussinov model
is not the only model for meson decay. An alternative approach to
string breaking, which does not involve the Schwinger pair production
process but rather the quantum fluctuations of the flux tube, was
developed by~\dcite{Kokoski:1985is}. Since their model is rather
different in spirit we will not discuss it here.

\subsection{Corrections due to masses and angular momentum}

The model described above is one of the main ingredients of the
so-called Lund fragmentation
model~\cite{Sjostrand:1982fn,Andersson:1983ia,b_lund}, used for
prediction of meson shower and hadronisation events in
accelerators. Two simple improvements have been suggested in the
literature. The first one consists of taking into account the presence
of massive particles at the string endpoints. In this case the linear
relation~$M = T_{\text{eff}}\,L$ between the length and the mass of
the meson is modified. In the approximation of small quark masses one
finds~\cite{Ida:1977uy}
\begin{equation}
\label{e:GuptaRosenzweig1}
\frac{L}{M}=\frac{2}{\pi\,T_{\text{eff}}}- \frac{m_1+m_2}{2 T_{\text{eff}} M} 
+ {\cal O}\left(\frac{m_i^2}{M^2}\right)\,.
\end{equation}
This relation has been derived by many authors, see
e.g.~\cite{Gupta:1994tx,Wilczek:2004im}. \dcite{Gupta:1994tx} applied this relation
to decay rates, and concluded that for a decay width~$\Gamma$ which is
linear in the length,~$\Gamma/M$ is no longer a constant. It was shown
in~\cite{Gupta:1994tx} that the ratio increases with the increase
of~$M$ until it reaches its universal value for large~$M$, i.e.~for a
small ratio~$m_q/M$.

Furthermore, the model of~\dcite{Gupta:1994tx} incorporates the
centrifugal barrier that the quark has to pass in the tunnelling
process of the pair creation.  The WKB approximation, which for the
case of a single pair creation with no centrifugal barrier reproduces
exactly the exponential factor of~\eqref{decaypro}, now reads
\begin{equation}\label{centrifugal}
P\sim \exp\left(-2\int _0^{r_c}\!{\rm d}r\, \sqrt{ (E-V(r))^2 - m_{q}^2-\frac{l(l+1)}{r^2}}\right)\,,
\end{equation} 
where~$m_q$ is the mass of the quark created, $r_c$~is the turning
point and~$l$ is the angular momentum of the tunnelling quark.  If the
quark tunnels from the point at a distance~$R_q$ from the centre of mass
to a point at a distance~$R_q+r$, then the quark acquires the angular
momentum of the vaporised segment of the string which can easily be
calculated in the limit of small~$r$. When this expression is inserted
into~\eqref{centrifugal} one finds that the probability for a split of
the string takes the form
\begin{equation}
P\sim \exp\left[-\frac{\pi m_q^2}{2T_{\text{eff}}}
  \frac{\left(1+\frac{w}{6m(1-w^2 R_q^2)}\right)^2}{%
               \left(1-\frac{w^3 R_q}{2T_{\text{eff}}(1-w^2R_q^2)^2}\right)^{3/2}} \right]\,,
\end{equation}
which means that there is an extra suppression factor which is
position dependent.  This yields a preference of the string to decay
in a symmetric fashion, i.e.~in the middle. The main net effect of the
centrifugal potential is to increase the stability of the meson.
In~\cite{Gupta:1994tx} a comparison with experimental data was made,
indicating that corrections due the massive string end points lead to
better agreement than with the massless approximation of the basic
model~\cite{Casher:1978wy}. Due to the lack of high-precision data for
decay widths, the detailed structure of the exponent could, however,
not be tested.

\subsection{Decay of mesons in the new picture}
\label{s:newpicture}

Having summarised the Casher-Neuberger-Nussinov model for meson decay,
as well as the various improvements of it, the question is now whether
we can reproduce those phenomenological formulae from the
holographically dual description. We have recalled the description of
high-spin mesons in section~\ref{s:highspinmesons}. These are U-shaped
strings, with two vertical sections that connect to the two endpoints
which are on one or two flavour branes, and a horizontal part that
stretches along the infrared ``wall''. Kinematically, this closely mimics
the mesons of the Lund model because, a was shown
by~\dcite{Kruczenski:2004me}, the vertical parts of the U-shaped
string behave as two massive particles attached to the endpoints of
the string.  Classically, this system is stable.

Quantum mechanically, the meson configuration is unstable. One
distinguishes decay modes due to fluctuations of the string endpoints
and those associated with the splitting of the string.  The former
translates into the production of low-spin mesons, which will be
discussed in a forthcoming publication. The process of splitting
implies the presence of two high-spin mesons in the outgoing state.
In the present paper we will focus exclusively on the last channel,
leaving the other channel for future work. Before we go into the
details of the computation, we will first describe this decay channel
qualitatively and highlight some universal features which are
independent of the actual calculation.

In the general setup described in section~\ref{s:flavour_branes}, the
system is built from three types of flavour branes characterised by
their distance from the ``wall'', namely, light medium and heavy ($l$,~$m$,~$h$)
flavour branes, and correspondingly by six types of mesons.  For a
meson to decay into two mesons it has to split, in such a way that the
new endpoints also lie on a flavour brane. The decay probability thus
naturally consists of two separate factors: the probability of the
string to split at a given point, multiplied with the probability that
this given point is actually located at a flavour brane.

The probability of an open string to split has been studied a long
time ago for strings with Neumann boundary conditions in flat
space-time~\cite{Mitchell:1987th,Dai:1989cp,Wilkinson:1989tb}.
Assuming that these results are qualitatively correct also in a curved
background\footnote{This assumption has been used also in the context
  of strings in \adss~\cite{Peeters:2004pt,Peeters:2005pb}.}, the first factor
of the decay width is therefore under control. The results
of~\cite{Mitchell:1987th,Dai:1989cp,Wilkinson:1989tb} show that the
open string decay probability per unit length is constant, or
equivalently, that the decay width is linear in the length of the
string. This linear scaling with the length is also present in the
Casher-Neuberger-Nussinov model.

\begin{figure}[t]
\begin{center}
\psfrag{approx}{\smaller\hspace{-1em} fluctuate + split}
\psfrag{suppressed}{\smaller\hspace{-1.5em} fluctuate + split}
\psfrag{trivial}{\smaller\hspace{.5em} only split}
\includegraphics*[width=.7\textwidth]{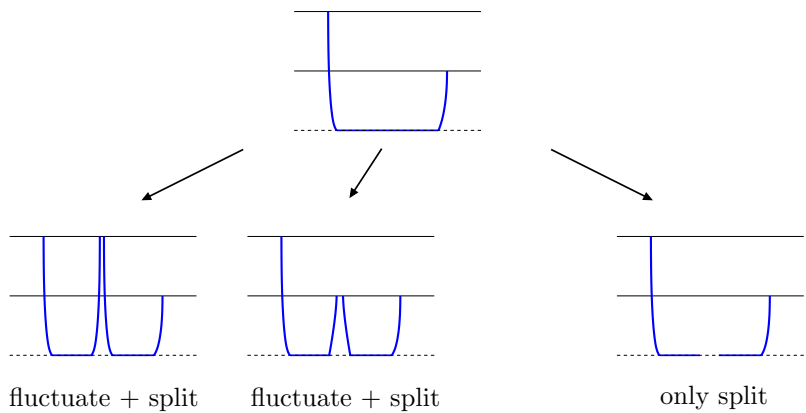}
\end{center}
\caption{The decay channels for a meson composed of one heavy and one
  intermediate-mass quark. When the newly produced quarks are massive,
  the computation of the decay width involves the computation of the
  probability that the string undergoes quantum fluctuations and
  touches a flavour brane. This is expected to lead to exponentially
  suppression (two figures on the left). Only when the new quarks are
  massless is the decay width given simply by the open string decay
  decay width.\label{f:channels}}
\end{figure}

The second factor is more complicated. If the string splits on the
infrared ``wall'', it corresponds to the creation of two massless quarks
at the new endpoints. This is clearly a very special situation. In the
general case, the string will first have to undergo quantum-mechanical
fluctuations, such that one or more points touch one of the flavour
branes associated to massive quarks. Schematically, any meson can
split into three kinds of mesons
\begin{equation}
(a, b) \rightarrow (a,c) + (c , b) 
\end{equation}
where the $a,b,c$ stands for $l, m$ and $h$.  In
figure~\ref{f:channels} we demonstrate the decay pattern of a meson
composed of one heavy and one intermediate-mass quark. Our goal will
be to show that this fluctuation probability is responsible for a
Gaussian suppression as a function of the mass of the newly created
quarks, just like in~\eqref{decaypro} of the Casher-Neuberger-Nussinov
model.

Technically, it is quite complicated to compute the fluctuation
probabilities, as it involves the quantum dynamics of the U-shaped
string in a curved background subject to non-trivial boundary
conditions. We will address this computation in detail in the next
section. However, several general features of meson decays can easily
be seen to be automatically satisfied, without further computations:
\begin{itemize}
\item
Due to the fact that a split involves only one flavour brane, it is
completely trivial in this geometrical picture that the decay obeys
the conservation of flavour symmetry. Due to the split, a new vertical
line coming into a certain flavour brane is necessarily followed by an
outgoing vertical line. If the former is assigned to be a
charge~$+$~then the latter has obviously a~$-$~charge and hence charge
is conserved.  If there is a set of $N_f$ flavour branes, then the
endpoints of the created pair of vertical strings are in the complex
conjugate representation of each other, and thus also the non-abelian
flavour symmetry is conserved.
\item
It is also clear that the pattern of decays depicted in
figure~\ref{f:channels} do not include processes that are suppressed
by the so-called \emph{Zweig rule}.  These suppressed decays,
described in figure~\ref{f:zweig}, involve the annihilation of the
original pair of quark anti-quark.  In our picture this involves
fluctuations that bring together the two endpoints. This is obviously
of higher order in~$g_s$ and hence further suppressed in the large-$N$
limit.
\begin{figure}[t]
\psfrag{q}{\smaller $q$}
\psfrag{qb}{\smaller $\bar q$}
\psfrag{q'}{\smaller $q'$}
\psfrag{qb'}{\smaller $\bar q'$}
\begin{center}
\includegraphics*[height=23ex]{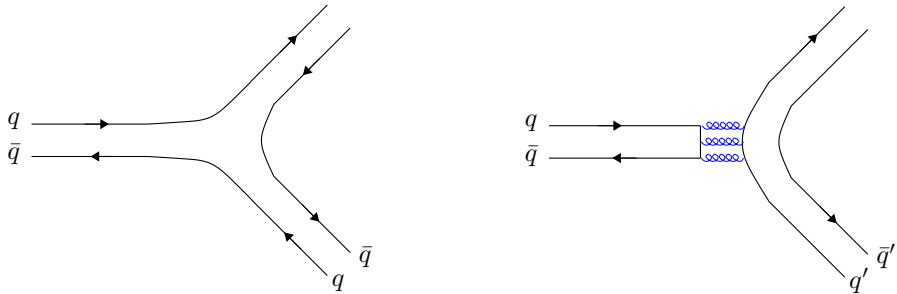}
\end{center}
\caption{The Zweig rule illustrated. The dominant decay channel for
  mesons is the process on the left, in which the original quarks are
  part of the mesons in the outgoing state. The process on the right,
  in which the quark and anti-quark which constitute the initial meson
  annihilate, is suppressed.\label{f:zweig}}
\end{figure}
 \end{itemize}

The decay of a meson is quite different from the decay of glueballs.
The reason is that an open string corresponding to a meson can
spontaneously split \emph{if and only if} the splitting point lies on
a flavour probe D6-brane. Thus, for the U-shaped string as in
figure~\ref{f:basicsetup}, no decays are possible which are as simple
as the decay process of closed strings. Instead, one has to take into
account the probability that the U-shaped string fluctuates and
touches the flavour brane. This is a true quantum-mechanical effect
and requires information beyond the probability of splitting a
string. In the next section, we will show that this effect can,
however, be computed in several approximations. We will thereby obtain
a prediction for the decay rate of mesons.

%%%%%%%%%%%%%%%%%%%%%%%%%%%%%%%%%%%%%%%%%%%%%%%%%%%%%%%%%%%%%%%%%%%%%%%

\section{Meson decay widths}
\label{s:decayrates}
\subsection{General remarks on wave functions, probabilities and widths}
\label{s:wavefunction}

In the previous sections we have reviewed the kinematical description
of mesons and glueballs in confining backgrounds, as well as the old
and the new ways to describe decay processes. We will now turn to a
quantitative analysis of the decay rates of mesons as computed using
the the gauge/string correspondence. We will see that this new way of
describing meson decay agrees also at the quantitative level with the
results of the old Casher-Neuberger-Nussinov
model~\cite{Casher:1978wy,Gupta:1994tx,b_lund}.  Before we go into the
details of this computation, we should first make some general
comments concerning the construction of the wave function and the
method to extract the decay width.

The general idea behind the construction of the string wave function
is the following.  One starts from the classical configuration of the
rotating U-shaped string. One then determines the spectrum of
fluctuations~$\delta X^M(\tau,\sigma)$ around this string
configuration.\footnote{Fermionic fluctuations are irrelevant for our
discussion and will be ignored throughout.} In order to be able to
quantise these fluctuations, they have to be written in decoupled
form, i.e.~in terms of normal coordinates. Generically, the normal
coordinates~$\{{\mathcal N}_n (X^M)\}$ are nontrivial functions of all
target space coordinates $X^M$, due to the fact that the target space
metric is curved.  Each mode ${\mathcal N}_n$ is described by its own
wave function $\Psi_n[{\mathcal N}_n]$, and the total wave function is
just a product of wave functions for the individual modes,
\begin{equation}
\label{simple}
\Psi\big[\{ {\mathcal N}_n \} \big] = \prod_n \Psi_n\big[{\mathcal N}_n (X^M)\big] \, .
\end{equation}
Analysing the system through the normal modes~${\mathcal N}_n$ is in
general not an easy task, because the space is curved and the normal
modes are thus hard to find.

Once the wave function is constructed, the first thing one would like
to extract is the probability that, due to quantum fluctuations, the
string touches the brane at one or more points. We will call this
probability~${\cal P}_{\text{fluct}}$ and it is formally given by
\begin{equation}
\label{fluct-prob}
{\mathcal P}_{\text{fluct}} = 
\int'_{\{ {\mathcal N}_n\}} \big|\, \Psi\big[\{{\mathcal N}_n\}\big] \, \big|^2\,,
\end{equation}
where the prime indicates that the integral is taken only over those
string configurations $\{ {\mathcal N}_n \}$ which satisfy the
condition
\begin{equation}
\label{e:abovebrane}
\max\big(U(\sigma)\big) \geq U_B\,.
\end{equation}
This is a complicated condition to take into account, because
$U(\sigma)$ is a linear combination of an infinite number of
modes. While the constraint is simple in terms of~$U(\sigma)$, it thus
becomes highly complicated in terms of the modes~${\cal N}_n$. The
probability~\eqref{fluct-prob} only measures how likely it is that the
string touches the brane, independent of the number of points that
touch the brane. Note that this is a dimensionless probability, not a
dimensionful decay width.

\begin{figure}[t]
\begin{center}
\psfrag{int}{{\Huge $\int\phantom{\vbox{\hbox{A}\hbox{A}\hbox{A}}}$}}
\psfrag{dx}{$\hspace{-1ex} K\big[\{{\cal N}_n\}\big]\,{\rm d}x$}
\psfrag{x}{$x$}
\psfrag{+}{$+$}
\psfrag{L}{$\!\!\!\!\approx\;\, 2L\;\times \kappa_{\text{max}}^{-1}$}
\includegraphics[width=.5\textwidth]{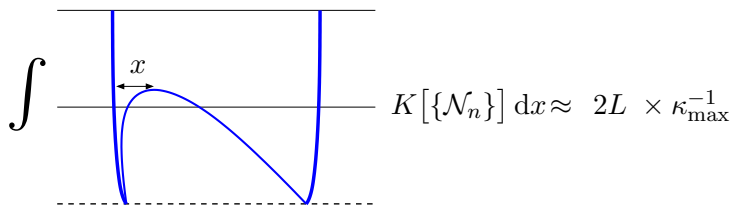}\hspace{6ex}
\end{center}
\caption{The approximation used to separate the $L$-dependent factor
  in the decay width from the dimensionless remainder. The integral
  over all configurations which touch the string at two points and
  have a maximum at~$U = U_B$ is, after taking into account the
  dimensionful measure factor~$K\big[\{{\cal N}_n\}\big]$,
  approximately equal to~$L$ times the volume of this subspace of
  configuration space.\label{f:approx1}}
\end{figure}

Let us now turn to the computation of the decay width itself. As we
have explained in section~\ref{s:newpicture}, we will assume that the
decay width of the mesonic string is approximately equal to the decay
width of an open string in flat space-time \emph{multiplied} with the
probability that the string actually touches the flavour brane. As we
have already mentioned (see also appendix~\ref{s:flatopen}), the decay
width of an open string in flat space-time has been shown to be linear
in the length~\cite{Mitchell:1987th,Dai:1989cp,Wilkinson:1989tb} (and,
for dimensional reasons, therefore inversely proportional to the
tension). We can thus define the ``decay width per unit length''
$\Gamma_{\text{open}}/L$, as well as a related, dimensionless,
\mbox{$L$-independent} quantity given by
\begin{equation}
\label{e:Psplitdef}
{\cal P}_{\text{split}} := \frac{1}{T_{\text{eff}}}\,\frac{\Gamma_{\text{open}}}{L}\,.
\end{equation}
In terms of this ``splitting probability'', the decay width of our
U-shaped string is now given by
\begin{equation}
\label{norm-prob}
\Gamma = T_{\text{eff}} {\cal P}_{\text{split}}\,\times\,
   \int'_{\{ {\mathcal N}_n\}} \big|\, \Psi\big[\{{\mathcal N}_n\}\big] \, \big|^2
   \, K\big[\{{\mathcal N}_n\}\big]  \, .
\end{equation}
The factor~$K\big[\{{\mathcal N}_n\}\big]$ is a measure factor with the
dimension of length. It measures, for a given string configuration,
the size of the segment(s) of the string which intersect(s) the
flavour brane.  We will not be very explicit about this factor. A
simple way to think about it is to consider e.g.~the subspace of
configurations with two intersection points for which the maximum
of~$U(\sigma)$ is fixed (see figure~\ref{f:approx1}). There is then
one direction in configuration space which effectively integrates over
all positions at which the string intersects the
brane. The~$K\big[\{{\mathcal N}_n\}\big]$ measures the infinitesimal size
of the intersection point(s) of the string with the brane. Provided
that the probabilities for the configurations in this integral are
more or less equal (for which we will find evidence in
section~\ref{s:stringbits}), we then obtain an overall factor~$L$ in
the decay width.  The overall factor~$L$ of course also arises
trivially for the zero mode fluctuation, where the string touches the
brane at all points at the same time.

We will not be able to compute~\eqref{norm-prob} as it stands, because
the factor~$K\big[\{{\cal N}_n\}\big]$ is too complicated to write down in
general. We will instead assume that any configuration is always part
of a one-parameter family of related configurations, which intersect
the brane at different points but otherwise have similar shape, as in
figure~\ref{f:approx1}. We will assume that the probability for all
these configurations is roughly the same, and that we can therefore
always split off a factor~$L$ from the integral. Typically this will
yield an upper bound on the decay width, because the configurations
with intersections in the middle typically have larger
probability. What remains is a dimensionless factor depending on the
position of the flavour brane. To be precise, we will compute the
right-hand side of
\begin{equation}
\label{e:approx2}
\Gamma = T_{\text{eff}}\,{\mathcal P}_{\text{split}} \,
   \int'_{\{ {\mathcal N}_n\}} \big|\, \Psi\big[\{{\mathcal N}_n\}\big] \,
   \big|^2 \, K\big[\{{\cal N}_n\}\big] \, <\, 
  T_{\text{eff}}\,{\mathcal P}_{\text{split}}\,L \,\kappa_{\text{max}}\, \int'_{\{ {\mathcal N}_n\}} 
  \big|\, \Psi\big[\{{\mathcal N}_n\}\big] \, \big|^2 
   \,,
\end{equation}
where~${\cal P}_{\text{fluct}}$ is given in~\eqref{fluct-prob} and
$\kappa_{\text{max}}$ is now dimensionless, arising from our approximation
(in case all configurations would be as in figure~\ref{f:approx1},
$\kappa_{\text{max}}$ would equal $1/\pi$). In the following, we will
therefore only be concerned with the computation of
\begin{equation}
\label{el}
\Gamma_{\text{approx}} 
 = \Big( T_{\text{eff}}\,{\mathcal P}_{\text{split}}\,\times\,
 L\,\times\, \kappa_{\text{max}}\Big)\,\times\, {\cal
   P}_{\text{fluct}}\,.
\end{equation}
In particular, we will not be concerned any more with the factor in
brackets, but focus solely on the dependence of the fluctuation
probability~${\cal P}_{\text{fluct}}$ on the position~$U_B$ of the
flavour brane.  This dependence on~$U_B$ translates to a dependence of
the decay width on the mass of the produced quarks.

Despite this simplification, the computation of the decay width is
still complicated, as the computation of~${\cal P}_{\text{fluct}}$
involves dealing with the curvature of the background and taking into
account the non-trivial constraint~\eqref{e:abovebrane}. In the
following sections, we will describe various approximation methods
which can be used to evaluate~${\cal P}_{\text{fluct}}$ and thereby
the decay width~$\Gamma$. A justification of these simplifications
will be obtained in section~\ref{s:stringbits}, where we compare the
continuum results with a numerical analysis using a string bit model.

\subsection{Explicit computation of the decay widths}

Let us first discuss the simplest type of approximation in which we
approximate the space-time near the ``wall'' with flat
space-time. There are two different configurations which may appear:
the light and the heavy mesons (see
section~\ref{s:highspinmesons}). Recall that in the case of a light
meson, i.e.~when the flavour brane associated to the initial quarks is
located on the ``wall'', the string endpoints satisfy Dirichlet
boundary conditions.  In the case of a heavy meson, the long vertical
parts of the string suppress, by their ``weight'', the fluctuations of
the part of the string near the end of the horizontal part. Therefore,
when viewed from the ``wall'', the string again looks like an open
string with Dirichlet boundary conditions. Hence, in both cases, one
can think about these strings, in first approximation, as being
attached to the ``brane'' which is located at the ``wall''.

Since the horizontal part of the string fluctuates near the ``wall'', it
experiences, at leading order approximation, a flat-space
geometry. Note though, that the Dirichlet boundary condition for
heavy mesons exist solely because of the vertical gravitational
potential. In this sense, the leading order flat-space approximation
refers only to the horizontal part of the string. As the amplitudes of
the fluctuations of the horizontal part of the string increase,
curvature effects set in and should be taken into account. Let us
first discuss the flat-space approximation.  In order to see when it
makes sense to use it, a useful intermediate step is to introduce a
coordinate
\begin{equation}
\eta^2 = \frac{U - U_\Lambda}{U_\Lambda}\,.
\end{equation}
The expansion of the metric~\eqref{D4} around $\eta=0$ yields, to
quadratic order~\cite{Bigazzi:2004ze},
\begin{multline}
\label{e:smalleta}
{\rm d}s^2 \sim \left( \frac{U_\Lambda}{R} \right)^{3/2} (1 + \frac{3}{2} \eta^2) (
\eta_{\mu\nu} {\rm d} X^\mu {\rm d}X^\nu)  + \frac{4}{3}  (R^3
U_\Lambda)^{1/2}  ({\rm d} \eta^2 +  \eta^2  {\rm d} \theta^2 )
\\[1ex] +
(R^3 U_\Lambda)^{1/2}  (1 + \frac{1}{2} \eta^2) {\rm d} \Omega_4^2 \, .
\end{multline}
We now want to quantised the fluctuations of the rigid rotating rod solution
which is sitting at the IR ``wall'' in the linearised metric~\eqref{e:smalleta}.
The solution is given by
\begin{equation}
\label{e:openstringrod}
T = L \tau\,, \quad
X^1 = L \sin\tau\sin\sigma\,,\quad
X^2 = L \cos\tau\sin\sigma\,,\quad U=U_\Lambda\,,
\end{equation}
where $\sigma\in [ -\pi/2, \pi/2 ]$. This same solution is valid both
for the light and heavy mesons, since there is no coupling between the
fluctuations along the ``wall'' and direction transverse to it.

There are two ways to quantise fluctuation around
(\ref{e:openstringrod}): using the Nambu-Goto or the Polyakov
formulation. The main idea and subtleties related to the presence of
the constraints are reviewed in the appendix.  The upshot of this
analysis is that fluctuations in the directions~$X^i$ transverse to
the plane in which the string rotates (i.e.~fluctuations in the
direction of the brane~$X^a$, the radial direction~$Z$ and in the
direction of internal sphere~$Y^m$) are massless in the flat space
approximation and become massive if the effects of curvature are taken
into account.  The fluctuations in the direction of the angular
coordinate in the plane where the string rotates are always massive,
with a sigma-dependent mass term.  As explained before, the
fluctuations in the direction of the ``wall'' are irrelevant for the
construction of the wave function. By expanding the Polyakov action
around the solution~\eqref{e:openstringrod} and keeping all terms
\emph{quadratic} in~$\eta$, we obtain the following action for the
fluctuations in the~$\eta$ and~$X^\mu$ directions,
\begin{multline}
\label{e:fluctaction}
S = 
\frac{1}{2\pi\alpha'} \left( \frac{U_\Lambda}{R} \right)^{3/2}  \int\!{\rm d}\tau{\rm d}\sigma\,  
 \frac{4}{3} \frac{R^3}{U_\Lambda} \left[ \left( \dot\eta^2 - {\eta'}^2\right)
 -\frac{8}{3} \, b \,\cos^2(\sigma)
\left( 1 + \frac{3}{2} \eta^2 \right) \,\right]  \\[1ex]
+ \left[ \left(1 + \frac{3}{2} \eta^2\right) \left(\delta \dot{X}^\mu \delta
  \dot{X}^\nu \eta_{\mu\nu}  - \delta X^{\mu '} \delta X^{\nu'}
  \eta_{\mu\nu}\right) \right] \, ,
\end{multline}
where we have introduced a dimensionless quantity,
\begin{equation}
\label{e:bdef}
b \equiv \frac{9}{16}\frac{L^2 U_\Lambda}{R^3} \, .
\end{equation}
We thus see that, unlike the linearised metric~\eqref{e:smalleta}, the
linearised action~\eqref{e:fluctaction} in addition to the small
parameter $\eta$ depends also on the extra parameter $b$, which
specifies what kind of string we are considering. Thus various
approximations will depend not only on the values of $\eta$, but also
$b$ and their relative ratio.  

To get a feeling for the meaning of the parameter~$b$, let us rewrite
it as
\begin{equation}
\label{brewrit}
b  =  \left( \frac{L}{L_\Lambda}\right)^2  = \frac{9}{16} 
 \frac{\alpha'}{{\mathcal R}^2 } \left( \frac{L^2}{\alpha'_{\text{eff}}} \right) \, , 
\end{equation}
where $L_\Lambda$ and ${\mathcal R}$ are defined in~\eqref{E0}
and~\eqref{E2} respectively. We see that if $ b\ll 1$, the
expression~\eqref{e:fluctaction} reduces to a flat-space action; this
case will be considered in the following section. As the size of the
fluctuations is increased, the string starts to see the curvature
effect; these additional corrections will be discussed in
section~\ref{s:curved}.  Note that the~\mbox{$b \ll 1$} regime is not
compatible with the decoupling of the Kaluza-Klein states: $b$~small
implies that the string is ``short'' enough to probe the extra compact
directions (i.e.~the energy of the string violates
condition~\eqref{E1}). However, semiclassical treatment of the string
still makes sense, as one can make the string macroscopic $(L^2 \gg
\alpha'_{\text{eff}}$).  This is possible as long as the supergravity
approximation is valid, i.e.~as long as~${\mathcal R}^2 \gg \alpha'$,
see~\eqref{brewrit}.

The situation is very different if $b \sim 1$.  A flat space limit is
\emph{not} possible in this case, regardless of the size of the
fluctuations.  Despite the fact that the string is fully localised
along the ``wall'', and probes the transverse directions only via small
fluctuations, if the length of the string along the ``wall'' is large
enough (in units of Kaluza-Klein radius $L_\Lambda$) the string will
always ``see'' curved transverse space.  The reason for this is the
potential term $b \eta^2$ in the action~\eqref{e:fluctaction} which
cannot be neglected for the low frequency modes (i.e. when
$\dot{\eta}\sim\eta'\sim \eta$).\footnote{The
action~\eqref{e:fluctaction} was obtained by linearising the Polyakov
action. In this approach, the constraints are easy to take care of at
leading order (see the appendix), but become more complicated at
higher orders. The linearisation of the Nambu-Goto form leads to a
more complicated action, but does not require any separate treatment
of the constraints. Therefore, studying the higher curvature effects
may be simpler in this approach.

The expansion the Nambu-Goto action in powers of~$\eta$ has the
schematic form
\begin{equation} 
\label{expq}
S = b ( q^2 + q^4 + \dots) + 
    (q^2 + q^4 + \dots) + 
    b^{-1} (q^4 + q^6 + \dots)+ 
   b^{-2} (q^6 + q^8 + \dots) + \ldots\, .
\end{equation}
where $q\sim \eta \sim \dot{\eta} \sim \eta'$. We thus see that the
expansion in~$\eta$ leads to a \emph{double expansion}, in~$q$ and~$b
q$. Hence, independent of whether $b\ll 1$ or $b \gg 1$, the parameter
$b q$ has to be much smaller than one for the semiclassical
expansion~\eqref{expq} to make sense.  In addition, the flat space
reduction makes sense if and~only~if
\begin{equation}
q^4/b  \ll q^2 \gg b q^2 \quad \Rightarrow  \quad   q^2 \ll b \ll 1 \, . 
\end{equation}}

\subsubsection{The flat space approximation}
\label{s:flat_no_wall}

As explained before, the action~\eqref{e:fluctaction} can be reduced
to the flat space action when~\mbox{$b\ll 1$}. Though this condition
violates the requirement that the Kaluza-Klein states decouple, this
is a generic problem of present models and we will ignore it in what
follows. Making the fluctuations~$\eta$ sufficiently small switches
off all curvature and allows us to write the metric~\eqref{e:smalleta}
in a conformally flat form, by introducing a new radial coordinate~$z$
as
\begin{equation}
\eta = \sqrt{\frac{3}{4}} U_\Lambda^{1/2} R^{-3/2}\, z\,.
\end{equation}
The metric then reduces to the simple form
\begin{equation}
\label{e:conf_flat_metric}
{\rm d}s^2 \sim \left( \frac{U_\Lambda}{R} \right)^{3/2}
\Big(\eta_{\mu\nu} {\rm d} X^\mu {\rm d}X^\nu
 + {\rm d}z^2 \Big) + (R^3U_\Lambda)^{1/2} {\rm d}\Omega^2_4\,.
\end{equation}
In this form it is immediately transparent that a string extended in
the~$X$ and~$\eta$ directions (but not in the four-sphere directions)
will be described by a flat-space string action, but with a string
tension which is given by~\eqref{e:Teff}.
The mass of a vertical string segment, stretching from the infrared
``wall'' to the flavour brane at $z=z_B$, is then simply given by 
\begin{equation}
\label{e:mq_zb}
m_q = T_{\text{eff}}\, z_B\,.
\end{equation}
The fluctuations in the direction of the angular coordinate in the
plane where the string rotates are massive, with a sigma-dependent mass
term. As explained before, the fluctuations in the direction of the
``wall'' are irrelevant for the construction of the wave function. By
expanding the Polyakov action around the
solution~\eqref{e:openstringrod}, we obtain the following action for
the fluctuations in the direction transverse to the ``wall''
\begin{equation}
\label{Sfluct}
S_{\text{fluct}} = \frac{L}{2 \pi \alpha'_{\text{eff}}} \int\!{\rm d} T {\rm d} \sigma\,  
\left[ - (\partial_T z)^2 + \frac{1}{L^2} (\partial_\sigma z)^2 + \ldots \right] \, .
\end{equation}
Here the dots refer to fluctuations in the directions along the
``wall''. Note that, by rotational symmetry, we can always align the
system such that fluctuations in the \mbox{$\theta$-direction}
decouple.  Taking into account the Dirichlet boundary conditions, the
fluctuations~$z(\sigma, \tau)$ can be written as
\begin{equation}
\label{ansatz}
z(\sigma,\tau) = \sum_{n>0} z_n \cos(n\sigma)\,.
\end{equation}
Using this expression in the action and integrating over the~$\sigma$
coordinate, the action for the fluctuations in the $z$-direction
reduces to
\begin{equation}
\label{lotLHO}
S_{\text{fluct}} = \frac{L}{2 \alpha'_{\text{eff}}} \int\!{\rm d} T\,
    \left[  \sum_{n>0}\left( - (\partial_T z_n)^2 +
		\frac{n^2}{L^2} 
  z_n^2 + \ldots \right)\right]\,.
\end{equation}
The main result which we deduce from this formula is that the system
is equivalent to an infinite number of uncoupled linear harmonic
oscillators, with frequencies~$n/L$ and masses~$L/\alpha'_{\text{eff}}$.  Note that
the form of the action~\eqref{lotLHO} (i.e.~the values of the masses
and the frequencies of the linear harmonic oscillators) is gauge
dependent. Thus to make computations more transparent, we have
intentionally chosen the static gauge on the worldsheet, so that these
worldsheet masses and frequencies coincide with their target space
values.  Note however, that the relevant combination $m \omega$ which
appears in the wave function is a \emph{gauge invariant} quantity.

We now write the wave function in the factorised form
\begin{equation}
\Psi(\{z_n\},\{ x_n \} )= \Psi_{\text{long}}(\{ x_n \}) \times 
                          \Psi_{\text{sphere}}(\{ y_n \}) \times
                          \Psi_\theta(\{ \theta_n \}) \times
                          \Psi_{\text{trans}}(\{ z_n \}) \, .
\end{equation}
Because the fluctuations along the ``wall'' and in the compact directions
are irrelevant for the computation of the probability that the string
touches the flavour brane, one can simply ``integrate'' these out (and
thus effectively set $\Psi_{\text{long}}=\Psi_{\text{sphere}} 
=\Psi_\theta = 1$).
The relevant, transverse part of the wave function is now given by
\begin{equation}
\label{simple2}
\Psi[\{z_n \}] = \prod_{n=1}^\infty \Psi_{0}(z_n)\,,
\end{equation}
where the wave functions for the individual modes are given by
\begin{equation}
\label{wavefnf}
\Psi_0(z_n) = \left(\frac{n}{\pi \alpha'_{\text{eff}} } \right)^{1/4} \exp\left(-
  \frac{n}{2\alpha'_{\text{eff}}} \, z_n^2\right) \, , 
\end{equation}
where all coordinates $z_n$ are unconstrained (i.e.~run from
$(-\infty, +\infty)$).\footnote{By expanding the fluctuations in
modes, we have here found that the eigenfunctions of the energy
operator depend only on~$\alpha'_{\text{eff}}$ and \emph{not} on~$L$,
whereas the eigenvalues contain an overall~$1/L$ factor. This is
easiest to understand by looking at the~$L$-dependence of the target
space energy given in~\eqref{e:energy} of the appendix. It arises
solely through an overall multiplication of the world-sheet
Hamiltonian, and does not influence the width of the wave functions.}
Note that negative values of~$z$ correspond to a fluctuation of the
string in antipodal directions on the cigar (antipodal points
in~$\theta$); see also the discussion of the geometry
around~\eqref{D4}. However, since the flavour brane is a point in the
$\theta$-direction, the fluctuations in the negative $z$-direction
will not touch the flavour brane. Note that all oscillators are in the
ground state, as classically none of the modes are excited in the
$z$-direction.

We can obtain several estimates for the probability that the string
touches the flavour brane.  These are all easiest to obtain by turning
the problem upside down, and asking for bounds on the probability that
the string does \emph{not} touch the brane. A lower bound on this
probability is given by integrating over all values of the~$z_n$ for
which the sum of the~$z_n$ satisfy
\begin{equation}
\sum_{n > 0} \big|z_n\big| \leq z_B\,.
\end{equation}
(in words, this means that even if all modes add up constructively,
the total amplitude is still smaller than~$z_B$).  This leads to an \emph{upper} bound on the probability
for the string to touch the brane,
\begin{equation}
{\cal P}^{\text{max}}_{\text{fluct}} =  1 - \idotsint\limits_{\sum_{n>0} |z_n| \leq z_B}\,
\prod_{n=1}^\infty {\rm d}z_n \, \big| \Psi( \{ z_n \}) \big|^2  \, .
\end{equation}

\begin{figure}[t]
\begin{center}
\psfrag{Pmin}{\raisebox{1ex}{${\cal P}^{\text{min}}_{\text{fluct}}$}}
\psfrag{zb}{$\displaystyle\frac{z_B}{\sqrt{\alpha'_{\text{eff}}}}$}
\includegraphics*[width=.55\textwidth]{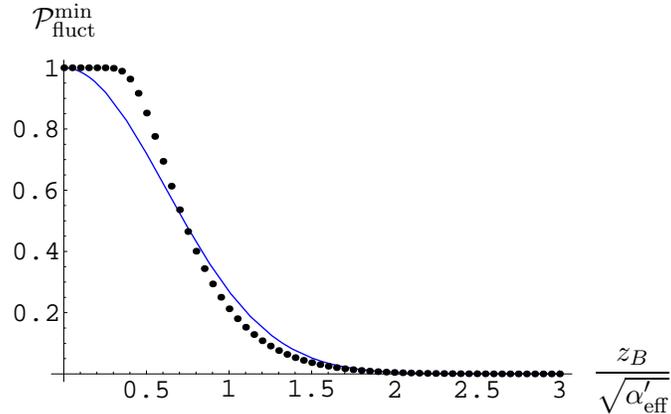}
\end{center}
\caption{Numerical evaluation of the lower bound on the
  probability~${\cal P}_{\text{min}}$ (i.e.~\protect\eqref{Pmin},
  for~$N\leq 1000$)
  that the string touches the flavour brane (black dots). Also
  displayed is a fit of the data to a Gaussian function~$\exp( -a
  z^2/\alpha'_{\text{eff}})$ (blue curve). The best fit gives $a \approx -1.3$.\label{f:fig1}}
\end{figure}

Another estimate can be constructed by considering the probability
that the string does \emph{not} touch the brane.  This probability can
be approximated by integrating over all values of the~$z_n$ for which
the amplitudes of the modes are smaller than~$z_B$ (this is likely to
overestimate the probability because it includes many configurations
for which the string actually touches the brane). From this, one
obtains a \emph{lower} estimate for the probability that the string
touches the brane,
\begin{equation}
\label{Pmin}
{\mathcal P}^{\text{min}}_{\text{fluct}} = 1- \lim_{N \rightarrow \infty} \,
\int_0^{z_B} \! {\rm d}z_1 \int_0^{z_B} \! {\rm d}z_2 \cdots
\int_0^{z_B} \! {\rm d}z_N \, \,  \big| \Psi( \{ z_n \}) \big|^2  \ .
\end{equation}
The integral~\eqref{Pmin} can be evaluated numerically, see
figure~\ref{f:fig1}. In order to see how well this fits the
Casher-Neuberger-Nussinov model, we have fitted the result to a
Gaussian. The result is a best fit given by
\begin{equation}
\label{expa}
{\mathcal P}^{\text{min}}_{\text{fluct}} \approx \exp\left( - 1.3\frac{z_B^2}{\alpha'_{\text{eff}}}\right) \, .
\end{equation}
From the plot we also see that, for small values of~$z_B$, the
deviation from the exponential suppression is more
prominent.

\begin{figure}[t]
\begin{center}
\vspace{2ex}
\psfrag{Pmin}{\raisebox{1ex}{${\cal P}^{\text{min}}_{\text{fluct}}$}}
\psfrag{zb}{$\displaystyle\frac{z_B}{\sqrt{\alpha'_{\text{eff}}}}$}
\includegraphics*[width=.55\textwidth]{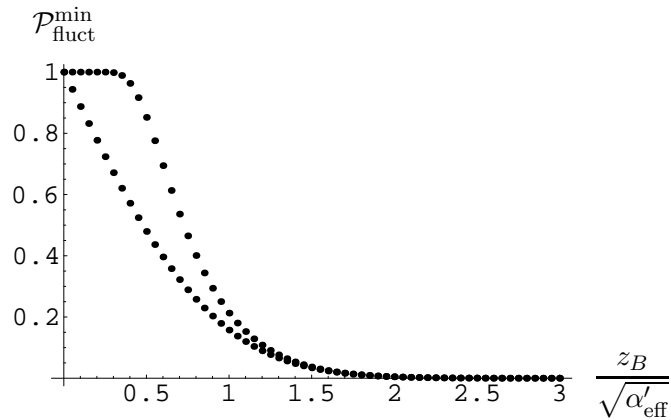}% all-vs-one.eps}
\end{center}
\caption{Comparison between the contribution to~${\cal
  P}^{\text{min}}_{\text{fluct}}$ of all modes~$N\leq 1000$ (upper `curve') vs.~only
  the lowest lying mode~$N=1$ (lower `curve'). The fit to a Gaussian
  is bad when only the lowest mode is taken into account.\label{f:compareN1N1000}}
\end{figure}

It is also illustrative to see the effect of the infinite number of
modes present in~\eqref{Pmin}.  We have therefore made a comparative
plot of ${\cal P}_{\text{min}}$ for~$N=1$ and for~$N\leq 1000$, see
figure~\ref{f:compareN1N1000}. 

As argued in section~\ref{s:wavefunction}, once the probability for
the fluctuation~${\mathcal P}_{\text{fluct}}$ is computed, the decay
width can be computed using~\eqref{el}. The $L$-dependent prefactor
in~\eqref{el} will be motivated further in
section~\eqref{s:stringbits} from a full numerical analysis of the
decay width using a string bit model. Combining~\eqref{el}
with~\eqref{expa} we obtain the following expression for the decay
width in the flat space approximation
\begin{equation}
\label{gammacont}
\Gamma_{\text{flat}} =  \Big( \text{const}.\times 
T_{\text{eff}}\,{\mathcal P}_{\text{split}}\,\times\,
 L \Big) \,\times\, \exp\left( - 1.3\frac{z_B^2}{\alpha'_{\text{eff}}}\right)  \, .
\end{equation}
Using~\eqref{e:mq_zb} we can compare this directly to the
Casher-Neuberger-Nussinov model.\footnote{We should emphasise that we
do \emph{not} claim that the probability is well-approximated by a
Gaussian for \emph{all} values of~$z_B$. This would imply a finite
value for the expectation value~$\langle z_B\rangle$, but the
expectation value in the ground state is known to
diverge~\cite{Karliner:1988hd}. Our results only show that for
relatively small~$z_B$ the probability is well-approximated by a
Gaussian. At large distances the decay has to go slower than Gaussian
to ensure divergence of~$\langle z_B\rangle$. However, the regime in
which the Casher-Neuberger-Nussinov model has been tested corresponds
to small~$z_B$. We thank Ofer Aharony for discussions on this issue.}
We will discuss this comparison in section~\ref{s:summary}. Let us
first analyse the effects that the curvature has on the decay width.

\subsubsection{Approximation using curved space}
\label{s:curved}

In this section we will discuss the effects of the curvature on the
mesons decay widths. In order to incorporate the leading effects of
curvature, we use the expansion of the D4-brane background~\eqref{D4}
around the ``wall'' at~$U = U_\Lambda$ as given in~\eqref{e:smalleta}. In
contrast to the situation discussed in section~\ref{s:flat_no_wall},
we will now consider the ``full'' metric~\eqref{e:smalleta} rather than
the truncated one~\eqref{e:conf_flat_metric}.

We now need to insert this metric in the string sigma model action and
expand the latter in small fluctuations around the classical,
rotating, U-shaped solution~\eqref{e:openstringrod}. The action for
the fluctuations in the~$\eta$ direction becomes
\begin{equation}
\label{e:etaction}
S = 
\frac{1}{2\pi\alpha'}\int\!{\rm d}\tau{\rm d}\sigma\,\left[
 \frac{4}{3}(R^3 U_\Lambda)^{1/2} \big( \dot\eta^2 - {\eta'}^2\big)
 -3\,L^2\,\cos^2(\sigma) \left(\frac{U_\Lambda}{R}\right)^{3/2}
 \eta^2\right]\,.
\end{equation}
Just as for the closed, folded string analysed
in~\cite{Bigazzi:2004ze}, we now find an effective equation of motion
for the fluctuation in the~$\eta$ direction which is given by
\begin{equation}
\label{e:mathieu}
\left[- \frac{{\rm d}^2}{{\rm d}\tau^2} +  \frac{{\rm d}^2}{{\rm d}\sigma^2} 
- \frac{9}{8}\frac{L^2\,U_\Lambda}{R^3}\,\big( 1 + \cos(2\sigma)\big) \right]
\eta(\tau,\sigma) = 0\,.
\end{equation}
We want to know the solutions to this equation subject to the
Dirichlet boundary conditions,
\begin{equation}
\label{e:bdycond}
\eta(\tau, -\frac{\pi}{2}) = \eta(\tau, \frac{\pi}{2}) = 0\,.
\end{equation}

Solutions of~\eqref{e:mathieu} and~\eqref{e:bdycond} only exist if
there is a non-zero and positive contribution to~\eqref{e:mathieu}
coming from the~\mbox{$-{\rm d}^2/{\rm d}\tau^2$} term. First, factorise
the solution according to
\begin{equation}
\label{e:factorise}
\eta(\tau,\sigma) = e^{i\omega \tau} \, f(\sigma)\,,
\end{equation}
with a real frequency~$\omega$. The resulting equation in~$\sigma$ is
the Mathieu equation~\cite{PandoZayas:2003yb,Bigazzi:2004ze}. The
solution which satisfies the boundary condition at the \emph{left} end
(i.e.~$\sigma=-\pi/2$) is given, up to an overall multiplicative
constant, by
\begin{multline}
\label{e:etasolD}
f(\sigma) = \\[1ex]
C\left( \omega^2 - b, \frac{b}{2}, -\frac{\pi}{2}\right)
      S\left(\omega^2-b,\frac{b}{2},\sigma\right)
     -S\left( \omega^2 - b, \frac{b}{2}, -\frac{\pi}{2}\right)
      C\left(\omega^2-b,\frac{b}{2},\sigma\right)\,.
\end{multline}
where~$C$ and $S$ are the Mathieu functions and~$b$ was defined
in~\eqref{e:bdef}.  We now need to tune~$\omega^2$ such that the
boundary condition at the \emph{right} end (i.e.$\sigma=\pi/2$) is
satisfied.

\begin{figure}[t]
\begin{center}
\psfrag{om}{$\omega$}
\psfrag{b}{$b=\frac{9}{16}\frac{L^2 U_\Lambda}{R^3}$}
\includegraphics[width=.6\textwidth]{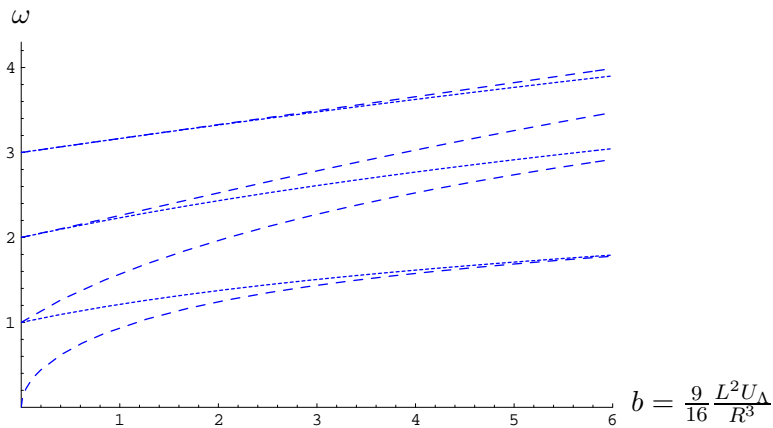}
\end{center}
\caption{The frequencies~$\omega$, given
  in~\protect\eqref{e:allfrequencies}, as a function of the
  parameter~$b$. These frequencies correspond to modes satisfying the
  equation of motion~\protect\eqref{e:mathieu} and the boundary
  conditions~\protect\eqref{e:bdycond}. Long dashes correspond
  to~$\omega^2_n = a_n(b/2)+b$ while short dashes correspond
  to~$\omega^2_n = b_n(b/2)+b$.  For $b=0$ and for
  $b\rightarrow\infty$, the spectrum is degenerate. For intermediate
  values of~$b$ there is level splitting.\label{f:frequencies}}
\end{figure}

This boundary condition at~$\sigma=\pi/2$ can be satisfied by making
use of the Mathieu characteristic functions~$a_n(q)$ and~$b_n(q)$,
which give the value of the first parameter of the even and odd
Mathieu functions respectively, such that they are periodic with
period~$2\pi n$.  We use the following properties of the Mathieu
functions,
\begin{equation}
\begin{aligned}
S( a_n(q), q, \pm\pi/2) = S( b_n(q), q, \pm\pi/2) &= 0  &&\quad \text{for even $n$,}\\[1ex]
C( a_n(q), q, \pm\pi/2) = C( b_n(q), q, \pm\pi/2) &= 0  &&\quad \text{for odd $n$.}
\end{aligned}
\end{equation}
These properties imply that for even~$n$, the second term
of~\eqref{e:etasolD} vanishes and the first one satisfies both
boundary conditions. For odd~$n$, the situation is reversed, and the
first term in~\eqref{e:etasolD} vanishes altogether while the second
term satisfies both boundary conditions. We thus see that the boundary
condition at~$\sigma=\pi/2$ is satisfied for any of the frequencies
\begin{equation}
\label{e:allfrequencies}
\begin{aligned}
\omega^2_n = a_n(b/2) + b \quad & \text{for $n\geq 0$}\,,\\[1ex]
\omega^2_n = b_n(b/2) + b \quad & \text{for $n> 0$}\,.
\end{aligned}
\end{equation} 
This spectrum has been plotted in figure~\ref{f:frequencies}. At
leading order these frequencies behave like~$n^2$ but there
are~$b$-dependent (and thus~$L$-dependent) corrections.

Knowing the frequencies of the modes, we can write down the
corresponding harmonic oscillator system. By writing the
action~\eqref{e:etaction} in terms of the target-space time~$T$ and
using the equation of motion to eliminate (after partial integration)
the ~$\eta''$ term, we find
\begin{equation}
S = \frac{L}{2\alpha'}\frac{4}{3}(R^3 U_\Lambda)^{1/2}
\int\!{\rm d}T\, \sum_n\left[
\left(\frac{{\rm d}\eta_n}{{\rm d}T}\right)^2
- \frac{\omega_n^2}{L^2} \eta_n^2 
\right]\,,
\end{equation}
where~$\eta_n$ denotes the amplitude of the $n$-th mode (observe that
there are actually two modes for all~$n\geq 1$).  The wave
function for the ground state of this harmonic oscillator behaves like
\begin{equation}
\label{effpsi}
\Psi[\eta_n] \sim \exp\left[ - \frac{2}{3\alpha'} (R^3U_\Lambda)^{1/2} 
    \big(a_n(b/2) + b\big)^{1/2}\, \eta_n^2 \right]\,.
\end{equation}
We thus see that the Gaussian suppression factor starts with an
$L$-independent term (as in flat space), but then receives corrections
which are $L$-dependent. In order to get a better feeling for the
physics stored in the wave function, let us rewrite the~$b$ parameter,
using the value valid for low-mass quarks~\eqref{EJ-low}, in terms of
gauge theory quantities. This leads to
\begin{equation}
b_{\text{light quarks}} = \frac{27}{4}\pi^2 \frac{J}{g^2_{\text{YM}} N}\,.
\end{equation}
The curvature corrections thus tend to \emph{suppress}, through the
exponential factor, the decay of higher-spin mesons (similar to the
effect of the centrifugal barrier of~\eqref{centrifugal}). One should,
however, keep in mind that \emph{both} in the old string model and in
our setup, the corrections due to finite quark
masses~\eqref{e:GuptaRosenzweig1} tend to \emph{enhance} the decay
as~$J$ increases. There are thus two competing effects. Unfortunately,
the experimental data of the decay of high-spin mesons into other
high-spin mesons is rather scarce. We will return to a comparison with
experiment in section~\ref{s:summary}.

\begin{figure}[t]
\begin{center}
\psfrag{s}{{\small $\sigma$}}
\psfrag{A}{}
\includegraphics[width=.5\textwidth]{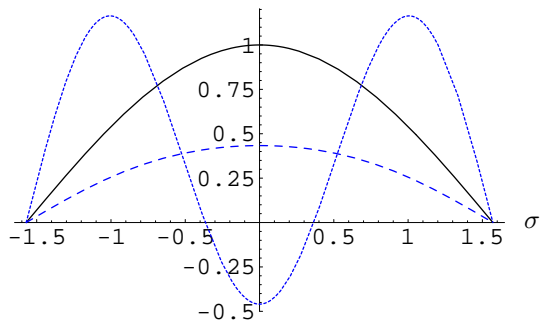}
\end{center}
\caption{The first excited modes (i.e.~corresponding to the
  frequencies for~$n=1$ in equation~\protect\eqref{e:allfrequencies}),
  with arbitrary normalisation. Dashing is as in
  figure~\protect\eqref{f:frequencies}. Also depicted is the mode
  which survives for~$b=0$, i.e.~in the absence of curvature (solid
  curve). \label{f:modesplitting}}
\end{figure}

%%%%%%%%%%%%%%%%%%%%%%%%%%%%%%%%%%%%%%%%%%%%%%%%%%%%%%%%%%%%%%%%%%%%

\subsection{Approximation using a string bit model in flat space}
\label{s:stringbits}

So far, we have used a continuum description to determine the decay
width. However, an alternative way to set up the computation is to use
a discrete approximation, where instead of a continuum string one uses
a set of beads and springs. This of course introduces a certain
approximation, but it has the advantage that the integration over the
right subset of configuration space becomes much more manageable. As a
result, we obtain an independent verification of the decay rates
obtained in the previous sections.

\begin{figure}[t]
\begin{center}
\psfrag{zb}{$z_B$}
\psfrag{-zb}{$-z_B$}
\psfrag{x1}{\small $x_1$}
\psfrag{a}{$a$}
\psfrag{L}{$L$}
\includegraphics*[width=.6\textwidth]{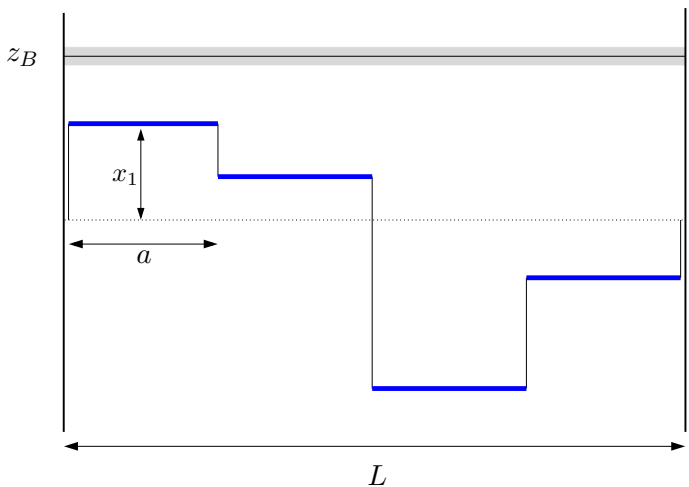}
\vspace{-2ex}
\end{center}
\caption{The discretised string, consisting of a number of horizontal
  rigid rod segments (blue) connected by vertical springs. Also
  depicted are the flavour brane at~$z=z_B$; this flavour brane has a
  finite width. \label{f:beadbox}}
\end{figure}

The situation we will consider is as in figure~\ref{f:beadbox}.  We
will again make the approximation in which the string is very close to
the ``wall'', so that the metric becomes flat, as
in~\eqref{e:conf_flat_metric}. The string tension of the string bit
model should thus be taken equal to~$T_{\text{eff}}$ given
in~\eqref{e:Teff}.

We now want to compute the probability that, when the system is in the
ground state, one or more beads are at the brane at~$x=z_B$.
In order to write down the wave function as a function of the
positions~$x_i$ of the beads, we have to go to normal coordinates in
which the equations of motion decouple. Denote the string tension
by~$T_{\text{eff}}$, the number of beads by~$N$, their individual
masses by~$M$ and the length of the system by~$L$, which satisfies~$L
= a\,(N+1)$. The action is given by
\begin{equation}
S = \frac{1}{2}\int\!{\rm d}t\, \left(\sum_{n=1}^N M\,\dot{x}_n^2
- \frac{T_{\text{eff}}}{a}\,\sum_{n=1}^{N+1} \big(x_n - x_{n-1}\big)^2\right)\,,
\end{equation}
where $x_{N+1}\equiv 0$. This action corresponds to taking Dirichlet
boundary conditions at the endpoints, i.e.~infinitely massive quarks
at the endpoints of the string. The normal modes and their frequencies
are then given by~\cite{b_georgiwaves}
\begin{equation}
\label{e:normalmodes}
y_m = \frac{1}{N+1}\sum_{n=1}^{N} \sin\left(\frac{m n \pi}{N+1}\right) x_n\,,\qquad
\omega_m^2 = \frac{4\,T_{\text{eff}}\,N(N+1)}{M_{\text{tot}}\, L}\, \sin^2\left( \frac{m \pi}{2 (N+1)}\right)\,,
\end{equation}
for which the inverse reads
\begin{equation}
x_n = 2\,\sum_{m=1}^N \sin\left(\frac{m n \pi}{N+1}\right) y_m\,.
\end{equation}
These expressions have been written in such a way that it is easy to
take the continuum limit~$N\rightarrow\infty$ while keeping~$L$, $T_{\text{eff}}$
and the total mass~$M_{\text{tot}}=N\,M$ fixed. In particular,
\begin{equation}
\lim_{N\rightarrow\infty} \omega_m^2 
 = \frac{m^2 \pi^2\, T_{\text{eff}}}{M_{\text{tot}}\,L}\,.
\end{equation}
In the relativistic limit~$M_{\text{tot}}=T_{\text{eff}}L$; we then get~$\omega_m^2
= m^2 \pi^2 / L^2$. The action then reads
\begin{equation}
\lim_{N\rightarrow\infty} S(M_{\text{tot}}=T_{\text{eff}}L) = 
T_{\text{eff}} \int\!{\rm d}t\int_0^L\!{\rm d}\sigma\,\Big[ \dot{x}(\sigma)^2 - {x'}(\sigma)^2\Big]\,.
\end{equation}
which will be useful for comparison later (note that~$\sigma$ is
normalised to run from~$0$ to~$L$).

The system is now decoupled and the action for the normal coordinates
is
\begin{equation}
S = (N+1) M\, \int\!{\rm d}t \sum_{m=1}^N \big( \dot{y}_m^2 - \omega_m^2 y_m^2\big)
\end{equation}
The wave function is a product of wave functions for the normal modes,
\begin{equation}
\Psi\big(\{y_1,y_2,\ldots\}\big) =
  \prod_{m=1}^N  \left(\frac{2(N+1)M \omega_m}{\pi}\right)^{1/4}
                 \exp\left( -(N+1)M \omega_m\,y_m^2 \right)\,.
\end{equation}
The wave function~$\Psi(\{x_1,x_2,\ldots\} )$ is now obtained simply
by inserting the normal modes \eqref{e:normalmodes}, which of course
results in a complicated exponential in terms of the~$x_n$. Note that
the width of the Gaussian behaves as
\begin{equation}
\lim_{N\rightarrow\infty} (N+1) M \omega_{m}
 = \lim_{N\rightarrow\infty} (N+1) \frac{T_{\text{eff}}L}{N}\frac{m\pi}{L}
 = T_{\text{eff}} \pi m\,.
\end{equation}
This expression depends linearly on~$T_{\text{eff}}$ and is independent of~$L$, in
agreement with the continuum analysis of section~\ref{s:wavefunction}.

The advantage of the discrete model is that we can now integrate the
square of the wave function over precisely the right subspace of
configuration space in order to determine the probability that the
string touches the brane (remember that this is the step which is hard
in the continuum, because in the continuum those boundary conditions
have to be rewritten as conditions on the normal modes). For each bead
position, we define the integration intervals corresponding to being
``at the brane'' and being ``elsewhere in space'' by
\begin{equation}
\label{e:bdydefs}
\begin{aligned}
I_{\text{brane}} &: \big[ -z_B - \Delta,\, -z_B \big]\; \cup \;
                    \big[ z_B,\, z_B + \Delta \big]\,,\\[1ex]
I_{\text{space}}  &: \big\langle -\infty,\, -z_B - \Delta \big]\; \cup\;
                    \big[-z_B,\, z_B\big]\; \cup\;
                    \big[ z_B+ \Delta,\, \infty\big\rangle\,.
\end{aligned}
\end{equation}
Here $\Delta$ is the width of the flavour brane, which of course has
to be taken equal to a finite value in order to be left with a finite
probability. The probability of finding a configuration which has,
e.g., one bead at the brane and all others away from it, is then given
by
\begin{multline}
\label{e:onebeadexample}
{\cal P}(\text{one bead at brane}) = \\[1ex]
 \sum_{i=1}^N  \int_{I_{\text{brane}}}\! {\rm d}x_i \;
         \prod_{k\not= i} \int_{I_{\text{space}}}\! {\rm d}x_k\;\;
         J\big(\{y_1,y_2,\ldots,y_N\},\{x_1,x_2,\ldots,x_N\}\big)\\[1ex]
         \times\Big|\Psi\big(\{x_1,x_2,\ldots,x_N\}\big)\Big|^2\,.
\end{multline}
The factor~$J$ is a Jacobian arising from the change of normal
coordinates to the original positions of the beads (and which is just
a constant since the transformation is linear, so it can be computed
by demanding that the integral of~$|\Psi|^2$ over the full~$x$-space is
equal to one).  Similar expressions exists for other subsets of
configuration space where more than one bead is located at the brane.

Let us now consider the simplest decay process, namely the one-meson
to two-meson process. The total decay width is a sum of decay widths
labelled by the number of beads which are at the brane,
\begin{equation}
\Gamma_{\text{meson} \rightarrow \text{2 mesons}}
  = \sum_{p} \Gamma^{(p)}_{\text{meson} \rightarrow \text{2 mesons}}\,.
\end{equation}
where the partial width~$\Gamma^{(p)}$ is given by
\begin{equation}
\label{e:bits_decay_width}
\Gamma^{(p)}_{\text{meson} \rightarrow \text{2 mesons}}
 = \sum_{\substack{\text{all configurations}\\\text{with $p$ beads at brane}}} p
 \,\cdot\, {\mathcal P}_{\text{configuration}}\,
 \cdot T_{\text{eff}}\,\, {\mathcal P}_{\text{split}}\,
 \cdot\, \text{length per bead}
\,.
\end{equation}
The factor~$p$ occurs because a configuration with~$p$ beads at the
brane can decay in~$p$ different ways into a two-string configuration.
The symbol ${\cal P}_{\text{split}}$ is the dimensionless coefficient
related to the decay width of open strings, as described
around~\eqref{e:Psplitdef}.  

In the discrete picture, it is easy to see that the decay width grows
linearly with the length of the string. Namely, consider the system
with~$T_{\text{eff}}$ fixed and $N$ fixed (and large in the continuum limit). The
total length (and thus the total mass) is now changed by varying the
spacing~$a$. The partial decay width~$\Gamma^{(1)}$, for instance, is
given by
\begin{multline}
\Gamma^{(1)}_{\text{meson} \rightarrow \text{2 mesons}} =
\sum_{i=1}^N 
      {\cal P}_{\text{bead $i$ at brane}}
\cdot T_{\text{eff}}\, {\cal P}_{\text{split}}
\cdot a \frac{N+1}{N}\\[1ex]
\approx 
      {\cal P}_{\text{one bead at brane}}
\cdot T_{\text{eff}}\, {\cal P}_{\text{split}}
\cdot L\,,
\end{multline}
where the last equality holds when the probability to sit at the brane
is approximately the same for all beads (recall, in this context, the
discussion in section~\ref{s:wavefunction}). In fact, as long as
$\sum_i {\cal P}_{\text{bead $i$ at brane}}$ scales linearly in~$N$,
one obtains a linear dependence of the decay width on~$L$. For the
partial width~$\Gamma^{(N)}$ the proportionality with~$L$ is in
actually trivial,
\begin{multline}
\Gamma^{(N)}_{\text{meson} \rightarrow \text{2 mesons}} =
      N
\cdot {\cal P}_{\text{all beads at brane}}
\cdot T_{\text{eff}}\, {\cal P}_{\text{split}}
\cdot a \frac{N+1}{N}\\[1ex]
=
      {\cal P}_{\text{all beads at brane}}
\cdot T_{\text{eff}}\, {\cal P}_{\text{split}}
\cdot L\,.
\end{multline}

\begin{figure}[t]
\begin{center}
\psfrag{P}{\hspace{-3.5em}\raisebox{2ex}{$\Gamma/(T_{\text{eff}} {\cal P}_{\text{split}} L)$}}
\psfrag{zb}{$\displaystyle\frac{z_B}{\sqrt{\alpha'_{\text{eff}}}}$}
\vspace{3ex}
\includegraphics*[width=.6\textwidth]{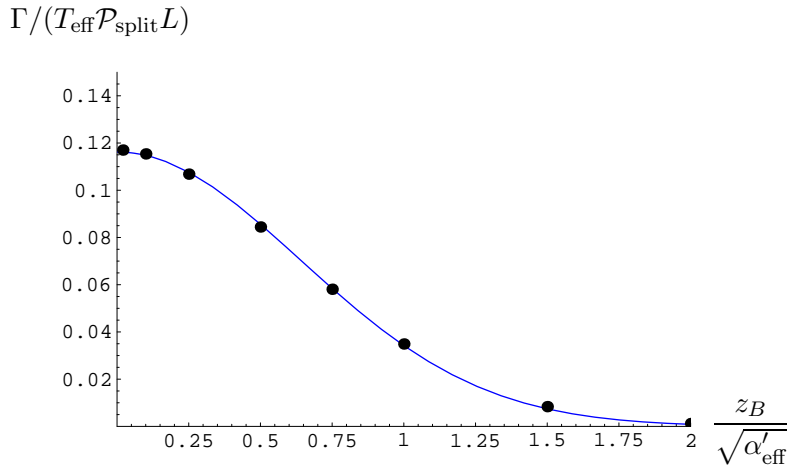}
\end{center}
\caption{Total decay width (divided by~$T_{\text{eff}}{\cal
  P}_{\text{split}}\,L$) for a six-bead system, with a brane width
  of~$0.1\,\sqrt{\alpha'_\text{eff}}$, as a function of the
  distance~$z_B$ of the IR~``wall'' to the flavour brane. Superposed is a
  best-fit Gaussian, with parameters $\Gamma = 0.12\,\exp( - 1.22\,
  z_B^2/\alpha'_{\text{eff}} )$. The string is allowed to split if a
  bead is at the brane, but the other beads are allowed to be anywhere
  (both below and above the brane).\label{f:sixblobgauss}}
\end{figure}
Let us now analyse the total decay width. Clearly, integrals of the
type~\eqref{e:onebeadexample} are complicated because they involve
exponentials which are not decoupled Gaussians (alternatively one
could of course integrate directly over the~$y_m$, as in the continuum
discussion, but then one has to deal with complicated boundary
conditions). However, one can certainly do these integrals numerically
using Monte-Carlo integration. 

With such a numerical integration process, we have obtained results
which are close to the ones which we obtained using approximation
methods in the continuum. An example of the decay width of a six-bead
system is given in figure~\ref{f:sixblobgauss}. By computing the decay
width for various values of~$N$ and extrapolating to large-$N$, one
finds that the decay width is well-approximated by
\begin{equation}
\label{e:bitsresult}
\Gamma_{\text{beads}} = \text{const.} \cdot \exp\left( - 1.0\,
\frac{z_B^2}{\alpha'_{\text{eff}}}\right) \cdot T_{\text{eff}}\,{\cal P}_{\text{split}}
 \cdot L\,.
\end{equation}
This extrapolation includes not just an extrapolation to large-$N$,
but also an extrapolation to small value of the brane
width.\footnote{After completion of this work we learned that a very
similar calculation was done, analytically, in the appendix
of~\cite{Kokoski:1985is}, though with an entirely different,
non-holographic underlying picture.}
\begin{figure}[t]
\begin{center}
\psfrag{width}{\hspace{-4em}\raisebox{2ex}{Gaussian exponent}}
\psfrag{N}{$N$}
\vspace{3ex}
\includegraphics*[width=.6\textwidth]{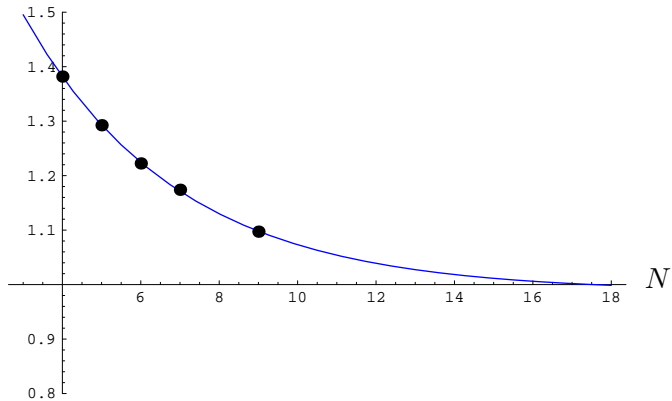}
\end{center}
\caption{The behaviour of the scale of the Gaussian shape of the decay
  width as a function of the number of beads~$N$. A best-fit
  exponential curve has been superposed, which suggest that the scale
  approaches one in the large-$N$ limit.\label{f:widthfit}}
\end{figure}

\begin{figure}[t]
\begin{center}
\vspace{3ex}
\psfrag{P}{\hspace{-3.5em}\raisebox{2ex}{$\Gamma/(T_{\text{eff}} {\cal P}_{\text{split}} L)$}}
\psfrag{zb}{$\displaystyle\frac{z_B}{\sqrt{\alpha'_{\text{eff}}}}$}
\includegraphics*[width=.6\textwidth]{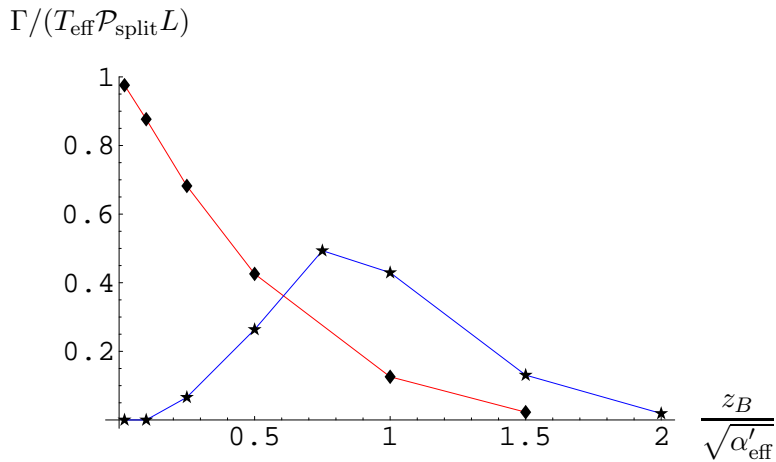}
\end{center}
\caption{Comparison of the effect of various incorrect prescriptions
  for the computation of the decay width (for five beads). When a bead
  is merely required to be at \emph{or} above the brane in order to be
  allowed to split, one obtains the red curve (diamonds). On the other
  hand, when the string is only allowed to split when \emph{none} of
  the beads are \emph{above} the brane, one obtains the blue curve
  (stars).
  \label{f:fivebitwrongbdy}}
\end{figure}

The match with the Gaussian curve is rather striking. It is perhaps
good to emphasise that this shape of the decay width is only obtained
when ``being at the brane'' and ``being somewhere else'' is defined as
in~\eqref{e:bdydefs}. If one allows the string to split not only when
a bead is at the brane, but \emph{also} when the bead is above the
brane, the decay width is qualitatively different. A different result
is also obtained if one disallows the string to split when any of the
beads is above the brane. See figure~\ref{f:fivebitwrongbdy} for
details.

The result~\eqref{e:bitsresult} is to be compared with the result of
the Casher-Neuberger-Nussinov model, given
by~\eqref{e:singledecay}. By using expression~\eqref{e:mq_zb} for the
mass of the produced quarks in terms of the distance between the ``wall''
and the flavour brane, we see that the decay width in their model,
when translated into string theory variables, is given by
\begin{equation}
\label{e:Lundresult}
\Gamma_{\text{CNN}} = \text{const.}\cdot 
\exp\left( - \frac{1}{4}\frac{z_B^2}{\alpha'_{\text{eff}}} \right)\cdot T_{\text{eff}}\,{\cal P}_{\text{split}}
 \cdot L\,.
\end{equation}
The exponents of~\eqref{e:bitsresult} and~\eqref{e:Lundresult} do not
agree. However, we should perhaps not be too surprised about this,
given the fact that it is not entirely clear yet whether the mass
appearing in~\eqref{mq} is constituent, current or something in
between.

Finally, let us remark once more that the results obtained with the
string bit model support the assumptions made in
section~\ref{s:wavefunction}, in particular those which lead to the
conclusion that the decay width is linear in the length of the string.

\section{Summary of the results and comparison with experiment}
\label{s:summary}

In this section we will summarise the results obtained throughout the
paper, and compare them with the available experimental data.  Since
our construction is very generic (i.e.~it does not crucially rely on
the details of the geometry dual to QCD) we focus on the
\emph{qualitative} features of the decay width as they follow from
our model, rather than on exact numbers.

In this paper we have studied decays of two types of high-spin mesons:
those consisting of two heavy or two light quarks. These mesons were
modelled by classical, open U-shaped strings (see
figure~\ref{f:Ustring}) with long and short vertical parts,
respectively.  The main idea for the computation of meson decay widths
was to study the fluctuations of the horizontal part of the U-shaped
string in the direction transverse to the ``wall'', and from there deduce
the probability that the string touches the flavour probe brane (see
figure~\ref{f:basicsetup}). This probability was then multiplied with
the probability for the string to reconnect with the probe, using the
flat space results for string decay widths of~\cite{Dai:1989cp}.

In order to compute the probability for the transverse fluctuations we
had to construct the string wave function~\eqref{simple}, for which it
was necessary to perform a semi-classical quantisation of the string
fluctuations around the classical U-shaped configuration. Given that
the full confining background~\eqref{D4} is complicated, and that we
are anyhow mainly interested in infrared phenomena of the theory, we
have focused in all our computations only on the region near the
``wall''. 

The description of the U-shaped strings also simplifies in this
approximation. Both for mesons with heavy quarks and for mesons with
light quarks, the horizontal part of the U-shaped string can be
modelled as an open string with Dirichlet boundary conditions on the
``wall''. This is obvious for the mesons with light quarks (where the
flavour brane is on the ``wall''). For mesons with heavy quarks the
vertical parts of the string cannot easily fluctuate in the transverse
direction in the full geometry due to their weight, and thereby impose
Dirichlet boundary conditions on the endpoints of the horizontal
segment. Note though, that the Dirichlet boundary condition for heavy
mesons exist solely because of the vertical gravitational
potential. At leading order, the horizontal part of the string
undergoes fluctuations near the wall just like in flat space. As the
amplitude of the fluctuations is increased, the curvature effects can
be taken into account perturbatively.

Within the frame work of these approximations the decay
width~$\Gamma_{\text{flat}}$ for heavy, high-spin
mesons~\eqref{gammacont} was shown to exhibit a)~linear dependence on
the string length (i.e.~the mass of the meson) b)~exponential
suppression with the masses of the produced quarks c)~flavour
conservation, d)~suppression in the large-$N_c$ limit and e)~the Zweig
rule.  The third feature is automatically built into the setup, and
corresponds to the geometrical fact that when a string splits, it
produces two endpoints on the same brane. The suppression in the
large-$N_c$ limit follows from the fact that reconnection of the
string to the brane is an open string process, and thus is weighted
with a $g_s \sim 1/N_c$ factor.  The Zweig rule, which is not
automatic in the Lund model, is similarly a simple consequence of the
holographic description of mesons. Processes violating the Zweig rule
(see section~\ref{s:newpicture}) would involve simultaneous
annihilation of the endpoints of the open string and splitting of the
string in the middle, and are thus suppressed by extra powers of the
string coupling constant.

\begin{figure}[t]
\begin{center}
\psfrag{K892}{\small \raisebox{-2ex}{$K^*(892)$}}
\psfrag{K1430}{\small \raisebox{-1.5ex}{$K^*_2(1430)$}}
\psfrag{K1780}{\small\hspace{-6em} $K^*_3(1780)$}
\psfrag{K2045}{\small $K^*_4(2045)$}
\psfrag{K2380}{\small $K^*_5(2380)$}
\psfrag{GdM}{\small $\Gamma/M$}
\psfrag{M}{\small $M~(\text{GeV})$}
\includegraphics[width=.7\textwidth]{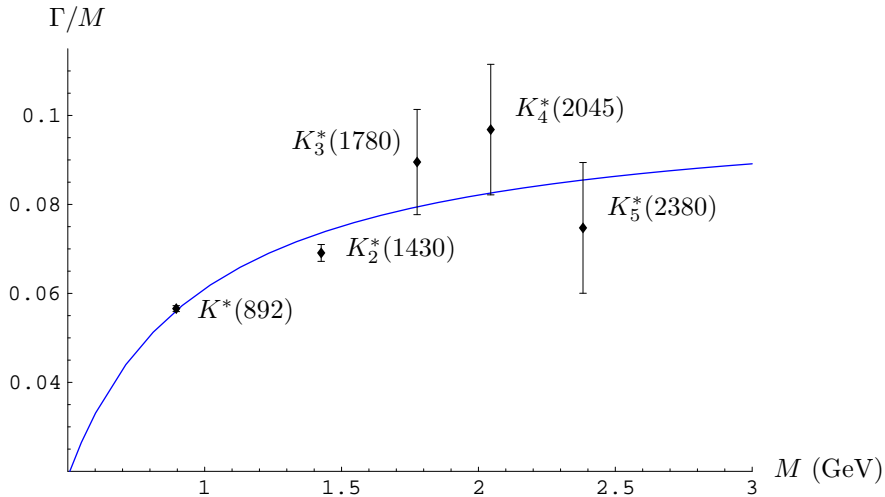}
\end{center}
\caption{The decay width divided by the mass of the mesons on
  the~$K^*$ trajectory, versus the mass of the
  states~\cite{PDBook}. These satisfy~$\Gamma/M=\text{const.}$ only
  very approximately. Also depicted is a fit based on the assumption
  of a constant width per unit \emph{length}, i.e.~\mbox{$\Gamma \sim L$},
  together with the finite quark mass relation between the length and
  the mass,~\protect\eqref{e:GuptaRosenzweig1}.\label{f:Kstar} Our
  model based on the gauge/string duality predicts a lower decay width
  at higher spin, but the data are too scarce to say anything
  conclusive about such corrections.}
\end{figure}

These first three of the features listed above are in very good
agreement with the experimentally tested Lund model
formula~\eqref{e:Lundresult} as computed in~\cite{Casher:1978wy}. In
relation to the exponential suppression of~$\Gamma$, we should in
particular note that the quark masses which figure in the Lund
result~\eqref{e:Lundresult} are the so-called ``constituent'' quark
masses, i.e.~masses which are of the order of hadronic masses (for
example~$\sim 300$ MeV for $u,d$ quarks, and $\sim 500$ MeV for the
$s$ quarks~\cite{Casher:1978wy}).  In our model the mass figuring in
the exponent is related to the mass of the vertical parts of the
string, and thus corresponds to the constituent mass rather than the
current algebra mass. The latter is related to the asymptotic
separation of the D4 and D6-branes (i.e.~to the parameter~$r_\infty$
in figure~\ref{f:D4D6}).\footnote{Note that even if~$r_\infty = 0$,
the constituent quark masses are non-zero, corresponding to the fact
that the non-flat profile of the D6-brane, as depicted in
figure~\ref{f:D4D6}, persists even in this case.}

The main differences between our results and those of the Lund model
are contained in the precise form of the exponent, in particular in
the dependence on the spin of the initial meson. Let us explain what
we mean here more precisely. However, as a general remark, note that
the fine structure of the exponent in~\eqref{e:Lundresult} has not
been extensively tested in experiments so far and in particular not
for the decay of high-spin mesons.

The quantity which can be extracted experimentally is~$\Gamma/M$,
where $\Gamma$ is the partial decay width and~$M$ the mass of the
initial meson. The Lund formula~\eqref{e:Lundresult} and our flat
space results both imply that the ratio~$\Gamma/L$ is \emph{constant}
on the same Regge trajectory, see~\eqref{e:bitsresult}
and~\eqref{e:Lundresult}.  Here~$L$ is the length of the horizontal
part of the U-shaped string, i.e.~the length of the flux tube in the
Lund model.  Thus, if one approximates mesonic Regge trajectories with
straight lines~(i.e~uses the condition $M= T_{\text{eff}} L$), then we
expect that the ratio $\Gamma/M$ is constant on a Regge trajectory.
The experimental data however, support this statement only very
approximately (see for example the decay widths for the
$K^*$~trajectory in figure~\ref{f:Kstar}).

There are at least two ways in which the observed deviations from
constant $\Gamma/M$ could be incorporated in the model. One way is by
incorporating the fact that in nature, Regge trajectories are not
straight lines and the relation between the meson mass and the
effective string ``length''~$L$ is modified, due to the presence of
massive quark endpoints~\eqref{e:GuptaRosenzweig1}.\footnote{This
simple mechanism cannot be used to explain deviations from the
straight line for the Pomeron trajectory (which corresponds to a
closed string).  In this case, the one-loop world-sheet corrections
do, however, modify the relation in the right
direction~\cite{PandoZayas:2003yb}.
Similar worldsheet quantum effects also modify the Regge trajectories
for mesons, but are much less relevant than the classical corrections
due to the quark masses.}
This \emph{kinematical} effect is automatically incorporated for our
U-shaped string, since the U-shaped string
satisfies~\eqref{e:GuptaRosenzweig1}. Note that this effect means that
for fixed energy, the string length is shorter, and hence the meson is
\emph{more} stable than with massless quarks.  As the spin of mesons
is increased, this effect is ``washed'' out, as the energy stored in
the string becomes much larger than the energy of the quarks. The
correction to~$\Gamma/M$ due to the finite quark masses indeed seems
to be in much better agreement with the experimental data, see for
instance the fit to the widths of the $K^*$~trajectory in figure~\ref{f:Kstar}.

The other reason for deviation of $\Gamma/M$ from a constant value
could be a modification to the exponential factor for the decay rates.
In the ``standard'' Lund model the ratio $\Gamma/M$ is ``blind'' to
the value of the spin of the mesons. The same is true for our model in
the flat space approximation, since the probabilities for the
fluctuations in the direction transverse to the ``wall'' were
\emph{independent} of the spin (i.e.~length) of the classical
string~\eqref{wavefnf}. However, both the Lund model and our setup
allow for an improvement of the exponential term to incorporate spin
effects.  In the Lund model the Schwinger pair production in the flux
tube is modified due to the presence of a centrifugal barrier
(see~\eqref{centrifugal} and the discussion around it). This effect
leads to a \emph{stabilisation} of mesons with the increase of their
spin.  In our setup, inclusion of curvature terms leads to the effect
that fluctuations in the transverse direction become \emph{dependent}
on the properties of the string around which they are excited
(see~\eqref{e:fluctaction}).  The leads to a twofold
effect~\eqref{effpsi}. Firstly, for a given meson, the lifetime is
increased as compared to the prediction in flat space, i.e.~as
compared to the simple Lund approximation~(it becomes harder to
fluctuate in the transverse direction due to the ``positive'' curving
of the space). Secondly, for a given trajectory, the mesons become
\emph{more stable} as their spin is increased. We would like to
emphasise that the direction in which these corrections act is very
generic, and basically independent of the particular model under
consideration.  Since both our model and the improved Lund model seem
to predict similar behaviour of the decay rates of high-spin mesons
(namely that they do not decay as fast as the naive Lund model
predicts), it would be extremely interesting to see whether
experimental data offers support for this behaviour. More precise data
to be collected in the future will hopefully allow for more detailed
tests of the exponential factors, and thus allow one to discriminate
between various models.

\newpage

\section*{Acknowledgements}

We would like to thank Johanna Erdmenger, Sergey Frolov, Vadim
Kaplu\-novsky, Shmuel Nussinov, Jan Plefka, Jorge Russo, Diana Vaman
and Shimon Yankielowicz for discussions. We would especially like to
thank Ofer Aharony for comments on the manuscript, and Stefano Kovacs
for participation at early stages of this project. J.S.~would like to
thank Stefan Theisen and the Max-Planck-Institut f\"ur
Gravitationsphysik for their hospitality. The work of J.S.~was
supported in part by the ISF grant.

\newpage
\appendix
\section{Appendix}
\subsection{Semiclassical quantisation of macroscopic strings}

In this appendix we discuss the general formalism required for
quantisation of fluctuations of the string around a given classical
solution. Most of this is based on appendix~A of~\dcite{Frolov:2002av}.

One starts from the Polyakov string action in a curved background,
gauge-fixed to a conformal worldsheet metric,
\begin{equation}
S = \frac{1}{2\pi\alpha'} \int\!{\rm d}\tau \int_0^{2\pi\sqrt{\alpha'}}\!{\rm d}\sigma
 \, G_{MN} \left[\dot X^M \dot X^N - X^M{}' X^N{}'\right]\,.
\end{equation}
In order to write down the lowest-energy state of the system, we have
to find an expression for operator corresponding to the space-time
energy of a string state. So we assume that the target-space metric
has been written proper-time form, $G_{0 i}=0$, and that all
components are independent of time so that it admits a time-like
Killing vector. The energy operator is then given by~$P_0$, or
\begin{equation}
E = P_0 = \frac{-1}{2\pi\alpha'} \int_0^{2\pi\sqrt{\alpha'}}\!{\rm d}\sigma\;G_{tt}\, \dot{t}\,.
\end{equation}
Using the Virasoro constraint, this expression can be converted to an
expression which involves the oscillators. The constraint reads
\begin{equation}
\label{e:Virasoro}
- G_{tt} \dot{t}^2 + G_{ij} \left[ 
   \dot{X}^i \dot{X}^j + {X'}^i {X'}^j \right] = 0\,.
\end{equation}
Now first expand the time coordinate around its classical value~$t=L
\tau + \tilde t$. The constraint~\eqref{e:Virasoro} then allows one to
derive the expression
\begin{equation}
G_{tt} {\dot t} = \frac{1}{2} L G_{tt} + \frac{1}{L} H(\tilde{t},X,Y,Z)\,.
\end{equation}
Here~$H(\tilde{t},X,Y,Z)$ is the world-sheet Hamiltonian density,
\begin{equation}
{\cal H}(\tilde{t},X,Y,Z) = - G_{tt} \Dot{\Tilde{t}}^2
+ G_{ij} \left[ \dot{X}^i \dot{X}^j + {X'}^i {X'}^j \right]\,.
\end{equation}
Now also expand the other fields around the classical solution, i.e.
\begin{equation}
X^i = X^i_0 + \tilde X^i\,,
\end{equation}
where~$X^i_0$ can for instance be the solution given
in~\eqref{e:openstringrod}. Upon integration over the~$\sigma$
coordinate, the terms linear in the fluctuations integrate to zero, so
that the result becomes
\begin{equation}
\label{e:energy}
E = \text{const.} + \frac{1}{L} \int_0^{2\pi\sqrt{\alpha'}}\!{\rm d}\sigma
\, {\cal H}(\tilde{T},\tilde{X},\tilde{Y},\tilde{Z})\,.
\end{equation}
This is the classical expression for which we want to write down the
corresponding quantum operator and the lowest-eigenvalue eigenstate.

\subsection{Decays of massive open strings in flat space-time}
\label{s:flatopen}

We briefly review here the decay of open bosonic strings and the
\mbox{type-I} superstrings via a split into two open strings. We
follow the papers \cite{Dai:1989cp,Wilkinson:1989tb}.

Intuitively, the string can split at any point of it and hence one
expects that the decay rate will be proportional to the length of the
string~$\Gamma\sim L$.  This property was indeed proved for the
bosonic critical open~\cite{Mitchell:1988qe,Dai:1989cp} and closed
string~\cite{Wilkinson:1989tb} as well as for the critical superstring
theory~\cite{Sundborg:1988ai,Amano:1988ht}.  The idea is to use the
optical theorem and compute the total decay rate by taking the
imaginary part of the self energy function. Whereas
in~\cite{Mitchell:1988qe,Wilkinson:1989tb} the annulus diagram
associated with the split and rejoining of an open string was
computed, the authors of~\cite{Dai:1989cp} use a trick and translate
the problem into that of disk amplitude with two closed string vertex
operators.  This is done by assuming one compact space dimension of
period~$L$ around which the initial and final string are wound.  So
the process is that of an incoming closed string that opens up and
then closes again to yield an outgoing closed string.

The corresponding amplitude takes the form
\begin{equation}
\label{splitamplitude}
i{\cal A} = \frac{iTN}{g^2} L \left [
\frac{\kappa}{2\pi\sqrt{L}}\right ]^2 \int_{|z|<1}\!{\rm d}^2z\, \langle:
e^{ip_0X(0)}::e^{-ip_0X(z)}:\rangle\,,
\end{equation} 
where $g$ is the coefficient of the open string tachyon operator,
$\kappa$ is the gravitational coupling, the normalisation
factor~$iTN/g^2$ is determined by calculating the amplitude of
a graviton to couple to two open string tachyons, the factor~$L$ comes
from the zero modes along the compact direction and
the~$1/\sqrt{L}$ factor for each vertex operator follows from
the normalisation of the centre of mass wave function of the string.
By using the operator product expansions of the left and right-moving
modes of the string one finds that the integrand
of~\eqref{splitamplitude} is given by
\begin{equation} 
\langle:e^{ip_0X(0)}::e^{-ip_0X(z)}:\rangle = |z\bar z|^{-2}( 1-z\bar z)^{-\gamma}\,,
\end{equation}
where $\gamma= \frac{L^2 T}{2\pi}-2$. Performing the integral, taking
the imaginary part of the amplitude and using Stirling's approximation
one finds that 
\begin{equation} 
\Imag {\cal A}= -\frac{TN\kappa^2}{2 g^2}\gamma 
\end{equation}
which means that in the large-$L$ limit the decay rate equals~\cite{Dai:1989cp} 
\begin{equation}
\label{split} \Gamma = - \Imag \delta m =
-\frac{1}{2m} \Imag {\cal A} = \frac{TN\kappa^2}{4 g^2E
}\gamma\rightarrow \frac{TN\kappa^2}{8 \pi g^2}L = \frac{g^2
T^{13}}{2^{26} \pi^{12}} N L \,.
\end{equation}
In the last step $\kappa$ was expressed in terms of~$g^2$ and~$T$
using either unitarity, careful treatment of the path integral
normalisation, or by factorisation of the annulus amplitude. The decay
rate is thus linear in the length of the string. Using the relation
between the length, the mass and the excitation level
\mbox{$M^2=n/\alpha'=L^2/\alpha'^2$} it is clear that
the decay rate is also linear in the mass of the string, and goes with
the square root of the excitation level. Eventually the decay rate has
units of~\mbox{$\text{time}^{-1}\sim M$} so we rewrite~\eqref{split}
in the following form
\begin{equation}
\Gamma = \frac{1}{\pi^{23}\sqrt{2^{45}}} M \,,
\end{equation} 
where, following \cite{Dai:1989cp}, the decay rate of~\eqref{split}
was multiplied by $\frac{16\pi}{g^2}\frac{1}{N\sqrt{(\pi T)^{25}}}$.

\begin{figure}[t]
\begin{center}
\includegraphics*[width=.25\textwidth]{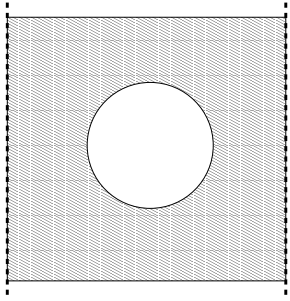}
\end{center}
\caption{The setup for the computation of the open string decay width
  as used by~\protect\dcite{Dai:1989cp}.}
\end{figure}

The calculations of Turok
et~al.~\cite{Mitchell:1988qe,Wilkinson:1989tb} include the evaluation
of the planar contribution to the self energy.  The self energy which
is related to the shift of the trajectory has threshold cuts along the
positive mass-squared axis, and the discontinuity across the cut gives
the decay rate. To determine the values of the decay rates of the
asymptotically high-mass states, the method of stationary phase was
used to compute the self energy integrals in the upper and lower half
planes.  It was shown that the decay rates are independent of the
endpoint charges, namely the decay of the different gauge
representation are the same.

The decay rate of~\cite{Mitchell:1988qe} does not agree in general
with the one computed by~\cite{Dai:1989cp}, as the former paper finds
a decay rate which is not (at leading order) linear in the length~$L$
of the string. However, the disagreement disappears in the critical
dimension (see the note added in proof
in~\cite{Mitchell:1988qe}). Since the two methods of calculations used
in~\cite{Dai:1989cp} and~\cite{Mitchell:1988qe,Wilkinson:1989tb} are
different, the fact that the results agree in the critical dimension
is significant. In particular it seems to imply that the impact of the
endpoints is not important and indeed the split is a local property.
If this conclusion is correct, it is more probable that the massive
endpoints of our open strings would not dramatically influence the
probability to split.

\newpage
\setlength{\bibsep}{3pt}

\begingroup\raggedright\endgroup

%\bibliographystyle{kasper}
%\bibliography{kasbib}
\end{document}